\documentclass[journal]{IEEEtran}
 
\usepackage{fontawesome5}
\usepackage{color}
\usepackage{graphicx, subfigure}
\usepackage{amsmath}
\usepackage{cite}
\usepackage{amsfonts}
\usepackage{multirow}
\usepackage{url}
\usepackage{threeparttable}
\usepackage{rotating}
\usepackage{hyperref}
\usepackage{booktabs}
\usepackage{makecell} 
\usepackage[many]{tcolorbox} 
\usepackage{colortbl}

\begin{document}

\title{LVM4CSI: Enabling Direct Application of Pre-Trained Large Vision Models for \\Wireless Channel Tasks} 
% \title{LVM4CSI: Directly Apply Large Vision Models to Wireless Channels}
\author{\normalsize {Jiajia~Guo, \IEEEmembership{\normalsize {Member,~IEEE}},
Peiwen Jiang,
Chao-Kai Wen, \IEEEmembership{\normalsize {Fellow,~IEEE}},
Shi Jin, \IEEEmembership{\normalsize {Fellow,~IEEE}}
and Jun Zhang, \IEEEmembership{\normalsize {Fellow,~IEEE}}\\  
Email: eejiajiaguo@ust.hk, eepwjiang@ust.hk, chaokai.wen@mail.nsysu.edu.tw, jinshi@seu.edu.cn, eejzhang@ust.hk
}
%\thanks{Jiajia Guo is with the Department of Electronic and Computer Engineering, The Hong Kong University of Science and Technology, Hong Kong, and was with National Mobile Communications Research Laboratory, Southeast University, Nanjing 210096, P. R. China (Email: jiajiaguo@seu.edu.cn).}
%\thanks{Peiwen Jiang and Jun Zhang are with the Department of Electronic and Computer Engineering, The Hong Kong University of Science and Technology, Hong Kong (Email: eepwjiang@ust.hk, eejzhang@ust.hk).}
%\thanks{Chao-Kai Wen is with the Institute of Communications Engineering, National Sun Yat-sen University, Kaohsiung 80424, Taiwan (e-mail: chaokai.wen@mail.nsysu.edu.tw).}
%\thanks{Shi Jin is with National Mobile Communications Research Laboratory, Southeast University, Nanjing 210096, P. R. China (Email:  jinshi@seu.edu.cn).}
}

\maketitle

\begin{abstract}
Accurate channel state information (CSI) is critical to the performance of wireless communication systems, especially with the increasing scale and complexity introduced by 5G and future 6G technologies. While artificial intelligence (AI) offers a promising approach to CSI acquisition and utilization, existing methods largely depend on task-specific neural networks (NNs) that require expert-driven design and large training datasets, limiting their generalizability and practicality. 
To address these challenges, we propose LVM4CSI, a general and efficient framework that leverages the structural similarity between CSI and computer vision (CV) data to directly apply large vision models (LVMs) pre-trained on extensive CV datasets to wireless tasks without any fine-tuning, in contrast to large language model-based methods that generally necessitate fine-tuning. LVM4CSI maps CSI tasks to analogous CV tasks, transforms complex-valued CSI into visual formats compatible with LVMs, and integrates lightweight trainable layers to adapt extracted features to specific communication objectives. 
We validate LVM4CSI through three representative case studies, including channel estimation, human activity recognition, and user localization. Results demonstrate that LVM4CSI achieves comparable or superior performance to task-specific NNs, including an improvement exceeding 9.61 dB in channel estimation and approximately 40\% reduction in localization error. Furthermore, it significantly reduces the number of trainable parameters and eliminates the need for task-specific NN design.
\end{abstract}

\begin{IEEEkeywords}
Large vision model, channel state information, channel estimation, human activity recognition, user localization.
\end{IEEEkeywords}

\IEEEpeerreviewmaketitle

\vspace{-0.5cm}
\section{Introduction}
\label{s1}

\IEEEPARstart{C}{hannel} state information (CSI) serves as the foundation of wireless communication systems. The performance gains from advanced techniques, such as massive multiple-input multiple-output (MIMO), orthogonal frequency-division multiplexing (OFDM), and integrated sensing and communication \cite{10763455,union2022future}, rely heavily on the accuracy of available CSI. To boost system performance, the scale of wireless systems, including the number of antennas and bandwidth, has progressively expanded from 1G to 5G and toward future 6G \cite{8808168}, but this scaling significantly increases CSI dimensions, thereby elevating the overhead and complexity of CSI acquisition and utilization. For instance, while 5G base stations (BSs) typically feature 32, 64, or even 128 antennas, this number is projected to rise to 512 in 6G \cite{3GPP250159}. Thus, the acquisition and utilization of high-dimensional CSI is becoming a significant challenge for the adoption of novel techniques in 6G and beyond. 

\subsection{Related Work}
\label{s111}
The deep integration of artificial intelligence (AI) with communications has emerged as a pivotal enabling technology for 6G \cite{10763455,union2022future,8808168,3GPP250159}, providing a promising solution to non-linear, high-dimensional, and complex communication problems \cite{10570412}. AI has been integrated into the wireless air interface, including channel estimation \cite{10368353}, channel prediction \cite{9210016}, CSI feedback \cite{guo2022overview}, CSI-based sensing \cite{10400499}, beamforming \cite{10144712}, among others, demonstrating exceptional performance and computational potential in addressing high-dimensional, complex communication challenges. Meanwhile, the industry has also actively researched AI-native air interfaces, encompassing CSI feedback \cite{guo2022overview}, beam management \cite{10627924}, and positioning \cite{10918333}, with significant advancements in the 3rd Generation Partnership Project (3GPP) Releases 18 and 19.

Despite significant progress, existing AI-based CSI acquisition and utilization methods still face several critical challenges, impeding their deployment in practical air interfaces.
\begin{itemize}
    \item {\bf Neural network design: }Although neural networks (NNs) can automatically learn from data, their architecture design heavily depends on expert knowledge and is both time- and computation-intensive. Numerous recent studies focus exclusively on integrating diverse NN architectures into communication tasks to enhance AI performance, while neglecting hardware compatibility.
    \item {\bf Large training data demand: }The superior performance of AI-driven communications relies heavily on extensive training data. However, data collection imposes significant overhead on communication and storage, and in certain scenarios, it is rendered infeasible by privacy regulations or system limitations.
    \item {\bf Generalization capacity: }A fundamental trade-off exists between achieving high performance and ensuring robust generalization in AI methodologies \cite{pmlryang20j}. AI models optimized for high performance on a specific dataset often overfit, leading to poor generalization across diverse datasets or dynamic communication environments.
\end{itemize}

Recently, large AI models (LAMs), such as large language models (LLMs) like ChatGPT and DeepSeek-R1 \cite{zhao2023survey}, have demonstrated significant potential in addressing the aforementioned challenges across domains such as natural language processing and computer vision (CV). In contrast to designing a specific NN model for a particular task or environment, LAMs can simultaneously handle multiple tasks or adapt to a single task across diverse environments, thereby eliminating the need for task- or environment-specific NN design \cite{guo2025prompt}. The LAM NN architecture is meticulously crafted, fully considering performance, complexity, hardware compatibility, and other factors, offering a potential solution to the first challenge. Moreover, as LAMs are pre-trained on extensive datasets, they can readily generalize to new environments \cite{alikhani2024large} and be fine-tuned with a relatively small amount of training data, thereby mitigating the latter two challenges.

\begin{figure}[t]
    \centering
    \includegraphics[width=0.5\textwidth]{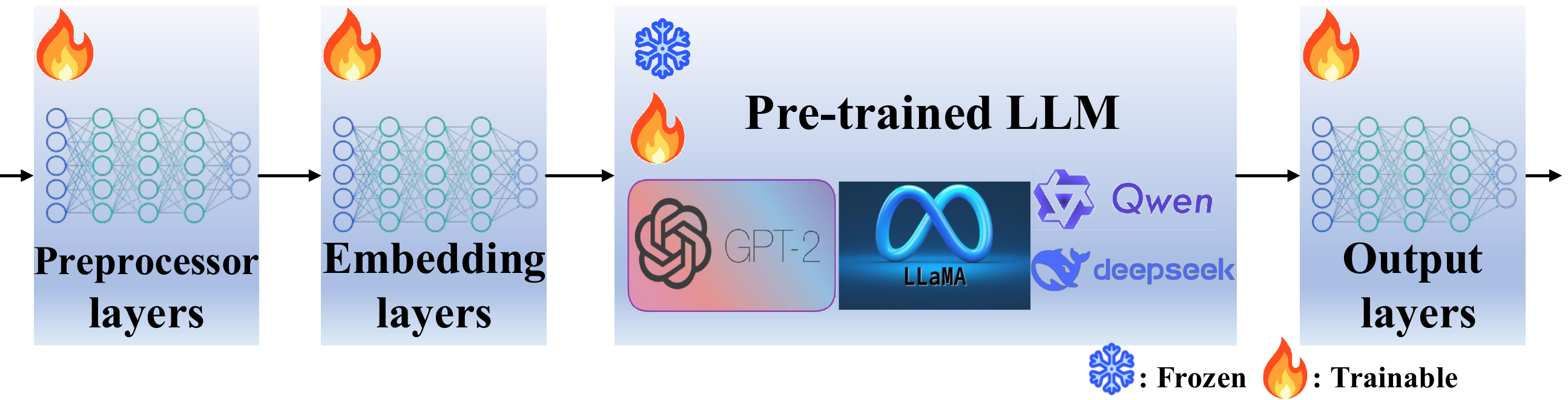} 
    \caption{\label{LLM4CP}A common framework of the existing LLM-based communication approaches \cite{10582829,cui2025exploring}. It adapts pre-trained LLMs with preprocessor, embedding, and output layers to align input wireless signals with the LLM's feature space, followed by fine-tuning with frozen parameters to boost accuracy.}
\end{figure}

Inspired by the significant potential of LAMs highlighted above, several preliminary studies have been proposed to address complex communication problems using LAMs, particularly LLMs. For instance, LLMs have been integrated into channel prediction \cite{10582829}, CSI feedback \cite{cui2025exploring}, beam prediction \cite{zheng2025beamllm}, port selection for fluid antennas \cite{zhang2025port}, signal detection \cite{zheng2024large}, and others. In these works, as shown in Fig. \ref{LLM4CP}, the pre-trained LLMs, such as GPT-2 and LLaMa 2, are embedded into the communication NNs to leverage their expressive power derived from extensive training data. For instance, the LLM4CP framework \cite{10582829} uses a pre-trained GPT-2 model to predict future downlink CSI sequences based on historical uplink CSI sequences. In particular, LLM4CP leverages the pre-trained GPT-2 by incorporating adaptive layers, including a preprocessor, an embedding layer, and an output module, to bridge the gap between the channel sequence and the LLM's feature space, followed by fine-tuning with partially frozen LLM parameters to enhance prediction accuracy.

The motivation for applying LLMs to wireless communications primarily stems from their ability to generalize across diverse tasks, suggesting potential applicability to wireless communication challenges~\cite{10582829,cui2025exploring,zheng2025beamllm,zhang2025port,zheng2024large,bu2022deep,10286020}. However, this motivation lacks a deep mechanistic understanding of the underlying principles, leaving uncertainty about the suitability of knowledge learned from natural language data for communication tasks. Although pre-trained LLMs are utilized, their parameters, as shown in Fig.~\ref{LLM4CP}, require partial fine-tuning to adapt to wireless tasks. Beyond adapting the output layer to align with the requirements of wireless tasks, a complex NN-based preprocessing module is employed to transform wireless data into a format compatible with the LLMs' input. These additional requirements highlight the fundamental differences between wireless and natural language data, and indicate that knowledge derived from natural language data cannot be directly applied to wireless tasks.

\begin{table*}[!ht]
\centering
\caption{Representative CV-inspired Methods for CSI Acquisition and Utilization}
\label{csi_cv_methods}
\begin{tabular}{lllll}
\toprule
\textbf{Application} & \textbf{Problem Addressed} & \textbf{Method} & \textbf{Reference}& \textbf{Publication Year}  \\
\midrule
\multirow{3}{*}{Channel Acquisition} 
  & Low-quality CSI denoising & DnCNN (Image Denoising) \cite{zhang2017beyond} & \cite{9127834}&2020 \\
  & Channel estimation with limited pilots & SRCNN (Image Super-Resolution) \cite{7115171} & \cite{10025776} &2023\\
  & Extracting path information from CSI & YOLO (Object Detection) \cite{7780460}& \cite{9120709, 10345484} &2020, 2024\\
\midrule
\multirow{5}{*}{CSI Utilization} 
  &User grouping & YOLO (Object Detection) \cite{7780460} & \cite{9495364} &2021\\
   & High-accuracy localization & CNN (Image Classification) \cite{ImageNetAlex} & \cite{8851471, 8468057} &2019, 2020\\
  & Gesture recognition  & VGG (Image Classification) \cite{vggnet}& \cite{bu2022deep}&2022 \\   
  & Human action understanding & U-Net (Medical Image Segmentation) \cite{10643318}& \cite{10286020}&2024 \\
  &Motion Source Recognition&ResNet (Image Classification) \cite{he2016deep} & \cite{10803907}&2025\\
\bottomrule
\end{tabular}
\end{table*}

\begin{figure*}[t]
 \centering
\subfigure[CSI image] {
 \label{CSIimage}
\includegraphics[width=0.22\linewidth]{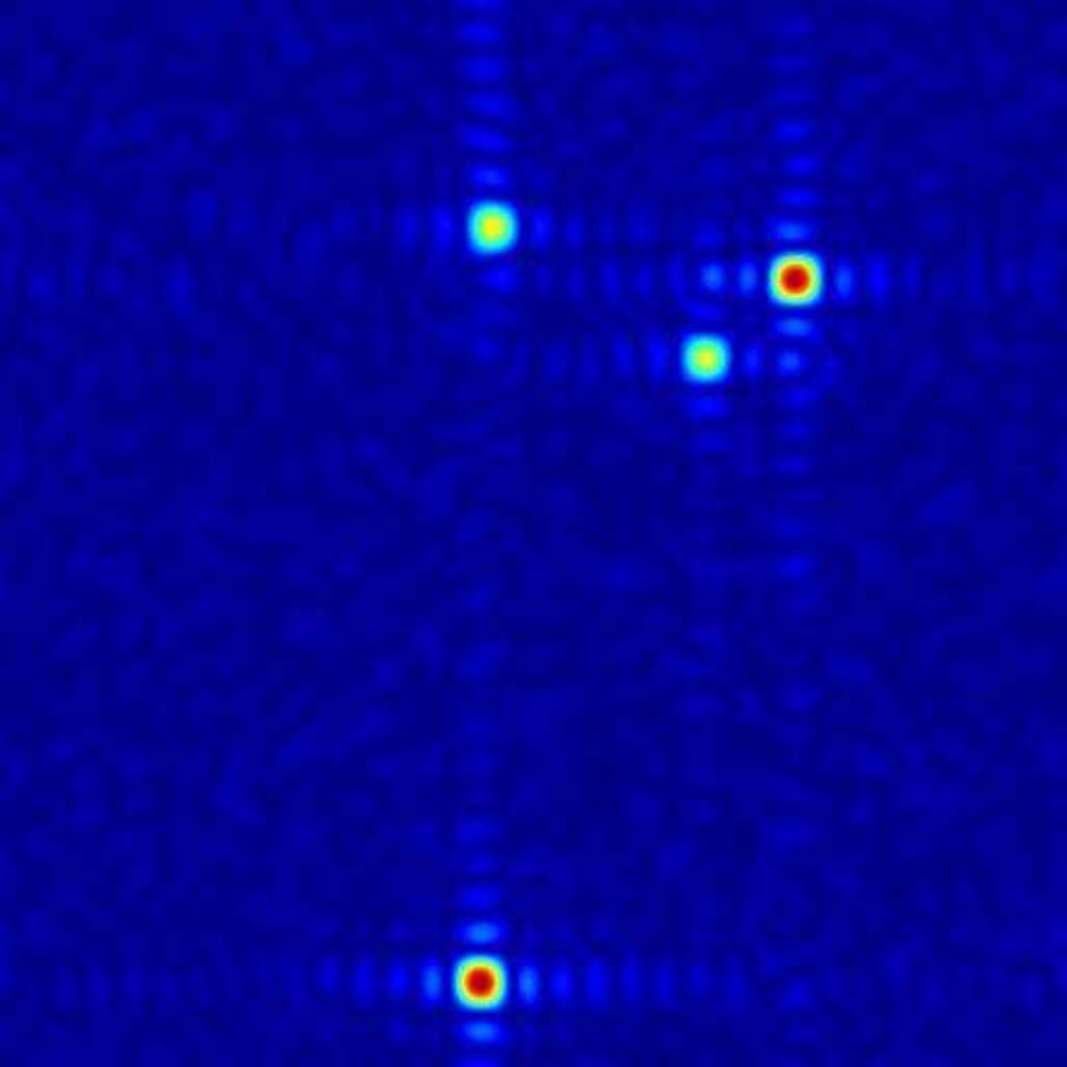}}
\subfigure[Starry sky image (ID: 1911 \cite{spaceyolodataset})] {
\label{star1911}
\includegraphics[width=0.22\linewidth]{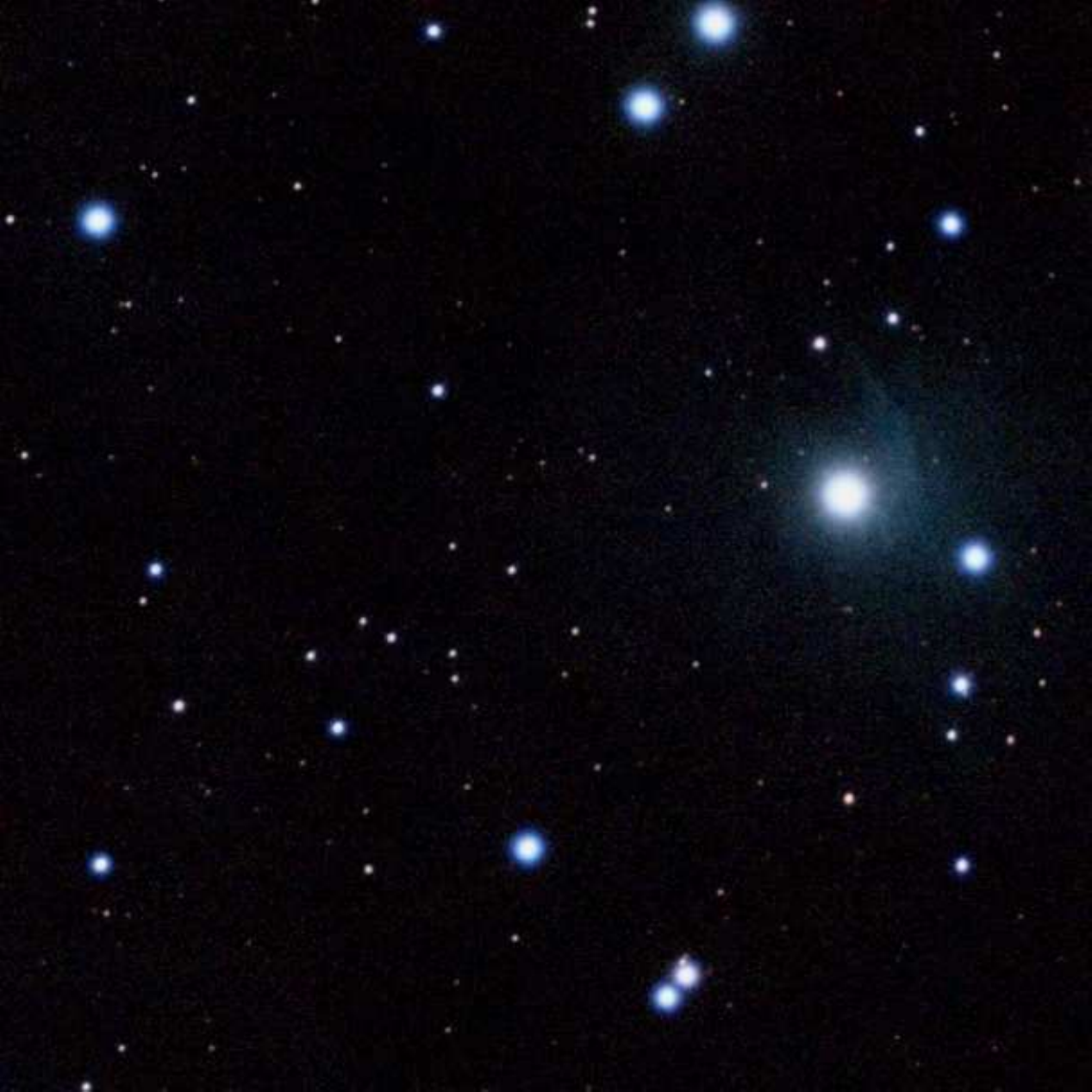}}
\subfigure[Detection result for CSI] {\label{CSIimageOD}
\includegraphics[width=0.22\linewidth]{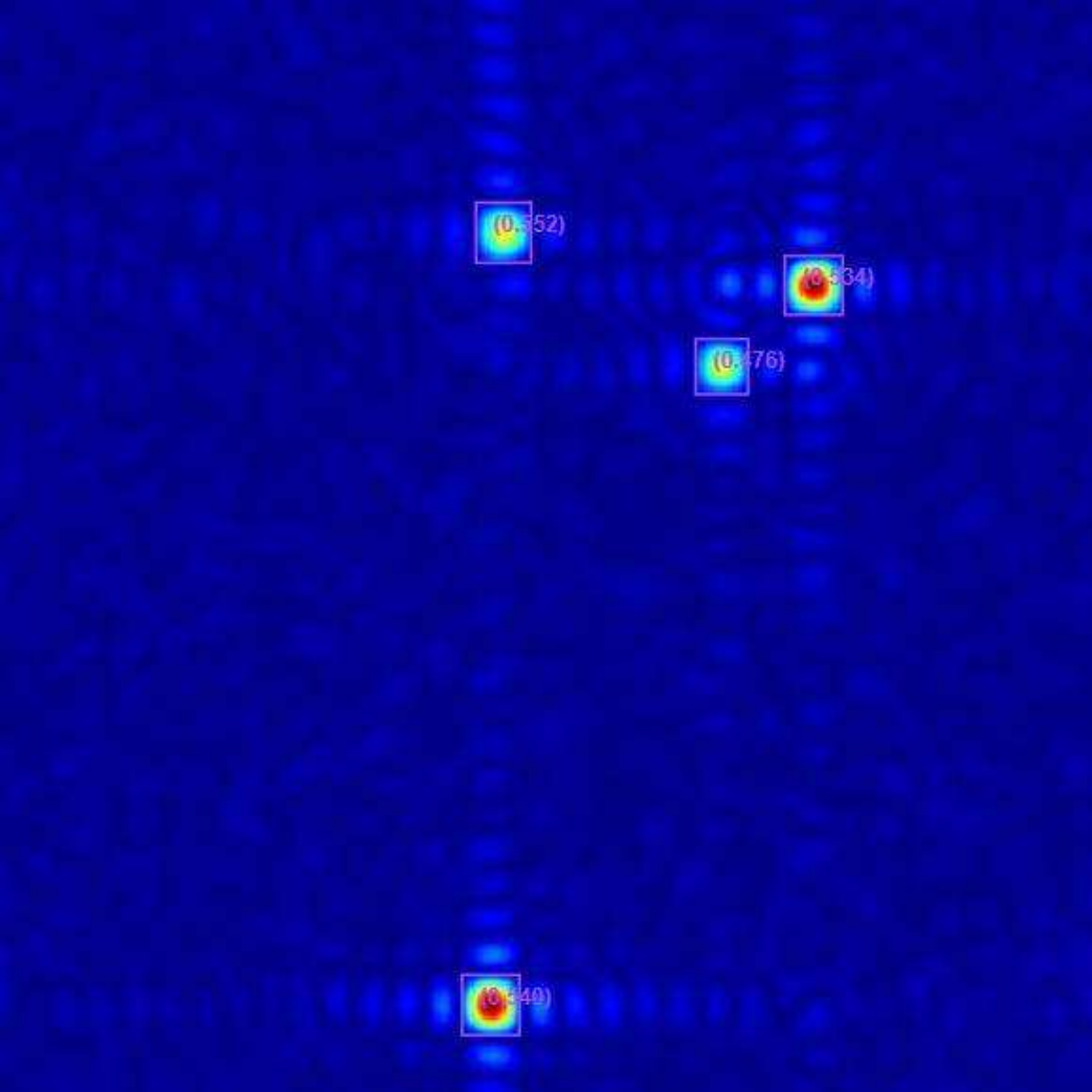}}
\subfigure[{Detection result for starry sky}] {
\label{star1911OD}
\includegraphics[width=0.22\linewidth]{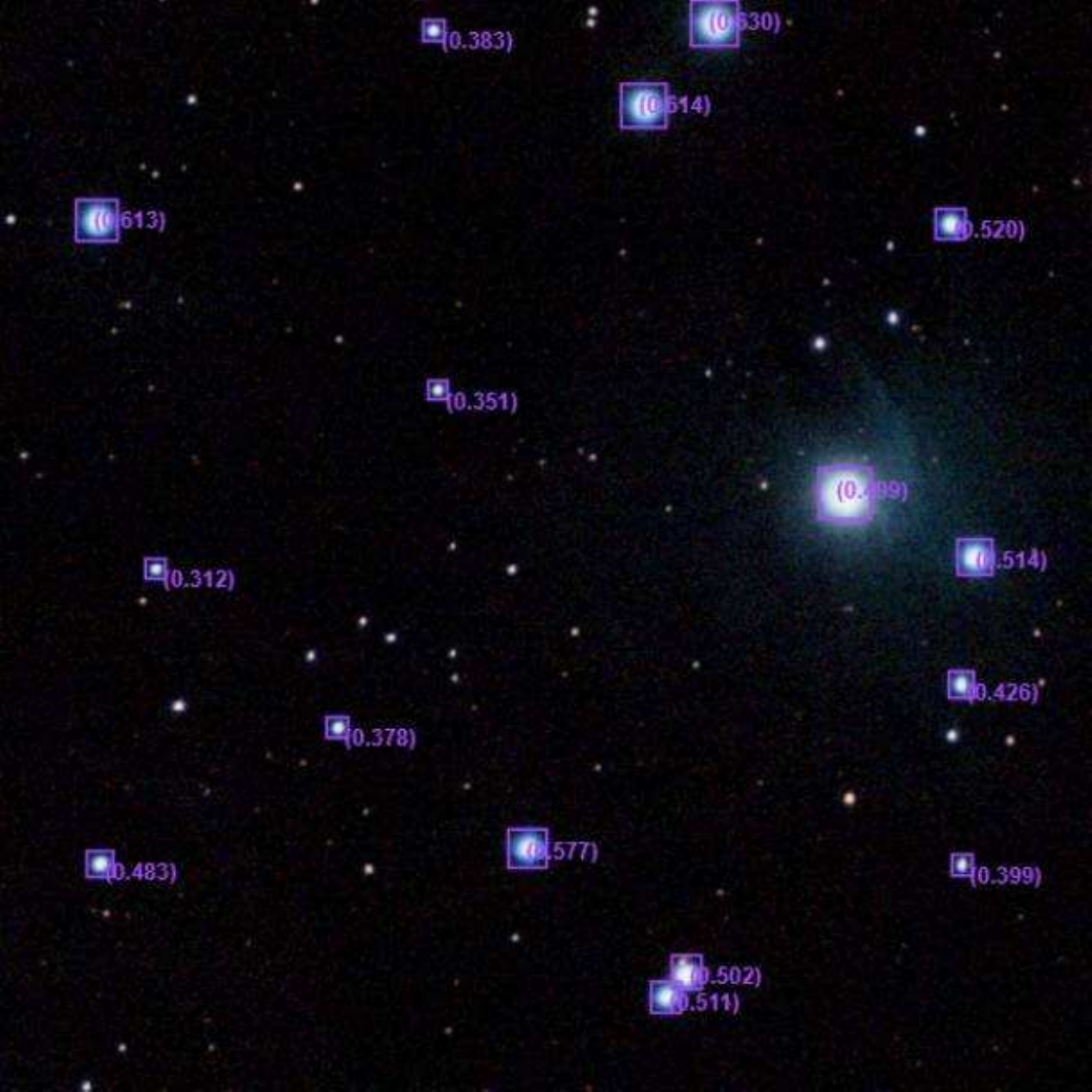}} 
\caption{Illustration of a CSI image \cite{9120709,10345484} and a starry sky image \cite{spaceyolodataset}. Despite representing entirely distinct modalities, they appear similar at first glance. The detected signal paths and stars are enclosed in purple boxes. Object detection performed by DINO-X \cite{ren2024dino} demonstrates significantly better performance in detecting signal paths in the CSI image compared to stars in the starry sky image. Note: Due to limitations of the object detection platform DINO-X, the object detection boxes in (c) and (d) are unclear.}
\label{CSIlikeImage}
\end{figure*}

In contrast to LLM-based methods that learn from natural language data, inspiration has been drawn from CV techniques for channel acquisition and utilization \cite{10570412}, as demonstrated in Table \ref{csi_cv_methods}. Specifically, leveraging the structural similarities between CSI and CV data, NN architectures originally developed for CV tasks have been extensively adapted for wireless communication applications, facilitating efficient processing and analysis. Ref. \cite{9127834} regards the estimated low-quality CSI as a two-channel noisy image and then uses a denoising convolutional NN (CNN), i.e., DnCNN, initially proposed for image denoising \cite{zhang2017beyond}, to improve the quality of estimated CSI. The authors in \cite{10025776} regard the channel estimation with limited pilots as an image super-resolution problem to recover and denoise the CSI matrix using super-resolution CNN, i.e., SRCNN \cite{7115171}. The object detection NN proposed for images, i.e., YOLO \cite{7780460}, is directly adopted in \cite{9120709,10345484} to extract path information from the CSI in the angular-delay domain.

Furthermore, techniques from CV have also been employed for CSI utilization. The object detection NN, YOLO \cite{7780460}, is employed for CSI-based user grouping in \cite{9495364}, where the statistical CSI of all users is projected onto a two-dimensional grid and used to partition the users into distinct groups. The authors in \cite{8851471,8468057} construct CSI ``images'' and utilize CNNs for image classification \cite{ImageNetAlex} to achieve high-accuracy localization. The VGG NN model \cite{vggnet}, originally trained for image classification, is utilized in \cite{bu2022deep} for CSI-based gesture recognition. The U-Net model, initially designed for medical image segmentation \cite{10643318}, is investigated in \cite{10286020} for CSI-based human action understanding. CSI-based motion source recognition is explored in \cite{10803907}, where a ResNet model, originally proposed for image classification \cite{he2016deep}, is utilized. By leveraging advanced NN architectures developed in the CV domain, the performance of the above CSI-related tasks is significantly improved.

\subsection{Motivation}

Owing to the fundamental differences between CSI and natural language data, LLM-based methods are not intuitively suitable for CSI-related tasks. In contrast, the structural similarities between CSI and CV data do exist. For example, Fig.~\ref{CSIimage} illustrates a CSI image in the angular-delay domain \cite{9120709,10345484}, characterized by a sparse pattern with four paths, while Fig.~\ref{star1911} depicts a starry sky image from \cite{spaceyolodataset}. Although these two images represent entirely distinct modalities, they appear similar at first glance. Consequently, knowledge derived from CV data could potentially be effectively applied to CSI-related tasks.
Although NN architectures originally developed for CV have been adapted in the aforementioned CV-based CSI studies, most of them train these NNs for specific tasks from scratch using CSI data. As NN knowledge is embedded in both their architectures and parameters \cite{WeightDistillation}, training CSI NNs from scratch without inheriting these parameters limits the full utilization of knowledge and insights gained from CV.

The reason NN parameters pre-trained on CV data are not commonly used is due to the general observation that wireless data and CV data constitute distinct modalities. This observation holds true when CV NNs are trained for specific tasks, such as handwritten digit recognition or cat and dog classification, which often limits their applicability to other data domains.
In contrast to task-specific NNs, large vision models (LVMs) are trained on a diverse array of CV data. For instance, classification NNs such as ConvNeXt \cite{Liu_2022_CVPR} are trained on extensive datasets like ImageNet with over 14 million images, while the object detection NN, DINO-X \cite{ren2024dino}, is trained on datasets exceeding 100 million image samples. These LVMs, pre-trained on extensive datasets, can be regarded as unified NN models with the potential to address cross-domain CV tasks, akin to LLMs in natural language processing.
For instance, Fig.~\ref{CSIimageOD} and Fig.~\ref{star1911OD} illustrate the object detection results for the two images obtained using DINO-X \cite{ren2024dino}. Although trained on CV images, the DINO-X model accurately identifies all four paths in the CSI image, whereas it detects stars with lower precision, as many small stars in the sky are challenging to capture.
Inspired by these promising results and astonished by the remarkable capabilities of LVMs, we raise the question:
\textit{Can we directly apply LVMs to solve CSI tasks in wireless communications?}

\subsection{Main Contributions of This Paper}
To address the above question, this study proposes LVM4CSI, a novel framework that directly applies LVMs, thoroughly trained on extensive CV datasets, to wireless channel acquisition and utilization. This simple approach allows us to harness the knowledge derived from CV data, in contrast to existing studies (e.g., those in Table \ref{csi_cv_methods}) that merely adopt NN architectures from CV or rely solely on their parameters as a starting point for fine-tuning.
The main contributions of this work are summarized as follows: 
\begin{itemize}
    \item {\bf LVM4CSI Framework:} We propose LVM4CSI, a novel framework for wireless channel acquisition and utilization. Particularly, observing that well-pre-trained, unified LVMs perform well on wireless channels, LVM4CSI applies these models, trained on a diverse array of CV datasets, to wireless channels without fine-tuning seamlessly leveraging their pre-acquired knowledge, in contrast to LLM-based methods that generally necessitate fine-tuning. 
    This framework eliminates the need for time-consuming, hardware-efficient, and high-performance NN design and training.
    
    \item {\bf Unified Workflow:} To facilitate the LVM4CSI framework, we present a detailed, unified workflow. Three core modules, including CSI-to-CV task translation, CSI-to-CV data transformation, and partial NN integration, are defined and thoroughly discussed. Through this workflow, LVMs can be employed to directly or indirectly address wireless channel tasks, such as efficient channel acquisition and highly accurate channel utilization.
    
    \item {\bf Case Studies:} Three case studies are conducted, namely channel estimation, CSI-based human activity recognition, and localization. The first focuses on channel acquisition, while the other two address channel utilization. Tailored NN frameworks are designed for these tasks. The pre-trained object detection LVM, DINO-X, is employed for high-quality path extraction in channel estimation, while pre-trained classification LVMs, such as ConvNeXt, are utilized to extract CSI features, enabling accurate human activity recognition and user localization.
    
    \item {\bf Simulation Validation:} Extensive simulation studies on diverse CSI datasets have been conducted to validate the effectiveness of the proposed LVM4CSI framework. By directly applying LVMs to wireless channels, this approach achieves performance comparable to non-AI methods and task-specific NNs designed and trained for the three cases mentioned earlier. Furthermore, the capacity of LVMs, often correlated with the number of NN parameters, significantly impacts the proposed method, necessitating a robust and unified LVM for optimal performance.
\end{itemize}
  
%\subsection{Paper Organization}
The rest of the paper is organized as follows. Section \ref{s2} introduces the system models of the considered channel-related tasks, i.e., channel estimation and CSI-based sensing. Then, Section \ref{s3} provides a comprehensive introduction to the proposed LVM4CSI framework and its workflow. Section \ref{s4} introduces the tailored NN frameworks for the three cases. Numerical results are provided and discussed in Section \ref{s5}. Finally, Section \ref{s6} concludes the paper and presents some future directions.

\section{System Models}
\label{s2}

\subsection{Signal and Channel Models}
We consider a single-cell massive MIMO system where the BS is equipped with a uniform linear array (ULA) consisting of $M \gg 1$ antennas, and the user has a single antenna. The antenna spacing at the BS and the signal wavelength are denoted by $d$ and $\lambda$, respectively. The system operates in time-division duplex (TDD) mode and employs $N$ subcarriers, with a subcarrier spacing of $\Delta f$.
The uplink transmission over the $n$-th subcarrier in this single-user massive MIMO system is given by 
\begin{equation}
\mathbf{y}_n = \mathbf{h}_n x_n + \mathbf{w}_n, \quad n = 1, 2, \ldots, N,
\end{equation}
where $\mathbf{h}_n \in \mathbb{C}^{M \times 1}$ represents the uplink channel, $x_n \in \mathbb{C}$ is the transmitted symbol, $\mathbf{w}_n \in \mathbb{C}^{M \times 1}$ denotes complex additive white Gaussian noise, and $\mathbf{y}_n \in \mathbb{C}^{M \times 1}$ is the received signal at the BS. The uplink CSI matrix in the spatial-frequency domain can be expressed as \cite{9120709} 
\begin{equation}
\mathbf{H} = \begin{bmatrix} \mathbf{h}_1 & \mathbf{h}_2 & \cdots & \mathbf{h}_N \end{bmatrix} \in \mathbb{C}^{M \times N},
\end{equation}
where $\mathbf{h}_n \in \mathbb{C}^{M \times 1}$ is the channel vector corresponding to the $n$-th subcarrier. The CSI matrix $\mathbf{H}$ plays a central role in wireless communications, as its quality and effective utilization significantly influence system performance in tasks such as transmission and sensing. 

The multipath channel between the BS and the user in the spatial-frequency domain can be modeled as 
\begin{equation}
\label{channelModel}
\mathbf{H} = \sum_{l=1}^L \alpha_l \mathbf{a}(\Theta_l) \mathbf{b}(T_l)^{\text{T}},
\end{equation}
where $L$ is the number of propagation paths, $\alpha_l \in \mathbb{C}$ is the complex gain of the $l$-th path, and $\mathbf{a}(\Theta_l) \in \mathbb{C}^{M \times 1}$ is the steering vector for the ULA, given by 
\begin{equation}
    \mathbf{a}(\Theta_l) = \left[ 1, e^{-j  2\pi  \Theta_l}, \ldots, e^{-j  2\pi  (M-1)  \Theta_l} \right]^{\text{T}},
\end{equation}
where $\Theta_l = \frac{d}{\lambda} \sin{\theta_l} \in [0, 1)$, and $\theta_l$ denotes the angle of the $l$-th path. The vector $\mathbf{b}(T_l) \in \mathbb{C}^{N \times 1}$ represents the delay-related phase components of the subcarriers and is expressed as 
\begin{equation}
    \mathbf{b}(T_l) = \left[ 1, e^{-j 2\pi T_l}, \ldots, e^{-j 2\pi (N-1) T_l} \right]^{\text{T}},
\end{equation}
where $T_l = \Delta f \tau_l \in [0, 1)$, and $\tau_l$ is the delay of the $l$-th path.

\subsection{Channel Estimation}
\label{s2B}

Considering the reciprocity between uplink and downlink channels in TDD mode, the downlink channel can be accurately inferred from the uplink. Therefore, this study focuses on uplink channel estimation. Assuming that all-one pilots are transmitted from the user to the BS, the received pilot signal at the BS, denoted by $\mathbf{Y} \in \mathbb{C}^{M \times N}$, is given by 
\begin{equation}
\label{pilotModel}
\mathbf{Y} = \sqrt{P} \mathbf{H} + \mathbf{W} = \sqrt{P} \sum_{l=1}^L \alpha_l \mathbf{a}(\Theta_l) \mathbf{b}(T_l)^{\text{H}} + \mathbf{W},
\end{equation}
where $P$ denotes the transmission power, and $\mathbf{W} \in \mathbb{C}^{M \times N}$ represents complex additive white Gaussian noise with zero mean and variance $\sigma^2$ across all $N$ subcarriers.
Since the number of paths $L$ is typically much smaller than the number of antennas $M$ ($L \ll M$), the channel estimation problem in (\ref{pilotModel}) can be formulated as extracting the parameter triplet $\{\alpha, \Theta, T\} = \{\alpha_l, \Theta_l, T_l \}_{l=1,\ldots,L}$ from the noisy received pilot signal $\mathbf{Y}$, and then reconstructing the channel using (\ref{channelModel}). 

The key challenge lies in accurately extracting the triplet $\{\alpha, \Theta, T\}$. Iterative methods such as the Newtonized Orthogonal Matching Pursuit (NOMP) algorithm \cite{7491265} are widely used in existing studies for channel parameter extraction. However, these methods exhibit high computational complexity and fail to meet the real-time requirements of practical systems. For example, the complexity of the NOMP algorithm is $\mathcal{O}(LMN \log_2(MN))$. When $M = 32$, $N = 32$, and $L = 2$, the algorithm's running time reaches up to 2.3 seconds \cite[Table II]{9120709}.
To address this issue, some studies \cite{9120709,10345484} have attempted to adopt advanced object detection techniques from CV to extract path information directly. This approach avoids the high complexity of iterative algorithms. The detailed process will be described in Section \ref{s41} and is therefore omitted here.

\subsection{CSI-based Sensing} 
Based on the observation that changes in the propagation environment significantly affect signal transmission, CSI can serve as an effective indicator for detecting environmental changes, that is, for performing sensing tasks \cite{10400499}. Therefore, we consider CSI-based sensing as a representative example of CSI utilization.
This sensing mechanism, which extracts information from a sequence of CSI measurements (or a single CSI instance), can be formulated as the following mapping: 
\begin{equation}
\{\mathbf{H}(t)\}_{t=1}^T \mapsto s_T,
\end{equation}
where $\mathbf{H}(t) \in \mathbb{C}^{M \times N}$ denotes the channel matrix at the $t$-th time slot, capturing environmental dynamics, and $s_T \in \mathbb{R}^D$ represents the inferred sensing output for the current time slot, with $D$ denoting the dimensionality of the sensed parameters.
 
In this study, we investigate two sensing tasks with different levels of granularity. The first task, human activity recognition, aims to classify human activities into discrete categories, such as falling, walking, running, and picking up objects \cite{8067693}. The second task, user localization, focuses on accurately estimating the user's continuous position.
Unlike human activity recognition, which classifies data into a small number of discrete labels, user localization is a regression task that predicts continuous positional values. This distinction makes user localization a more fine-grained and challenging sensing problem. 

\subsubsection{Human Activity Recognition}\label{HARmodel}
Since human activities occur over a sequence of time slots rather than instantaneously, they can only be inferred from a temporal series of observations. Therefore, the activity recognition task can be mathematically formulated as 
\begin{equation}
\label{harmodelF}
    \hat{y}_{\text{HAR}} = {\rm f}_{\text{HAR}}\left(\{\mathbf{H}(t)\}_{t=1}^T; {\bf \Theta}_{\text{HAR}}\right),
\end{equation}
where the NN model ${\rm f}_{\text{HAR}}(\cdot)$ outputs a probability distribution over $C$ activity classes after applying a softmax activation. The vector $\hat{y}_{\text{HAR}} \in \mathbb{R}^C$ represents the predicted class probabilities, and the final activity label is obtained via $\text{argmax}(\hat{y}_{\text{HAR}})$. Here, $\{\mathbf{H}(t)\}_{t=1}^T$ denotes the sequence of CSI matrices across $T$ time slots, and $\mathbf{\Theta}_{\text{HAR}}$ represents the parameters of the activity recognition network. The cross-entropy loss is used as the objective function for this classification task, defined as 
\begin{equation}\label{lossHAR}
    \mathcal{L}_{\text{HAR}} = - \frac{1}{S} \sum_{s=1}^S \sum_{c=1}^C y_{\text{true}}^{s,c} \log(\hat{y}_{\text{HAR} }^{s,c}),
\end{equation}
where $S$ is the number of samples in the batch, $y_{\text{true}}^{s,c}$ is the $c$-th element of the one-hot encoded ground-truth label for the $s$-th sample (equal to 1 if the $c$-th class is correct and 0 otherwise), and $\hat{y}_{\text{HAR}}^{s,c}$ is the predicted probability for the $c$-th class of the $s$-th sample. 

\subsubsection{User Localization}
The user’s position corresponds to a particular time slot and can therefore be directly inferred from a single CSI sample at that moment. Accordingly, the CSI-based localization task is formulated as 
\begin{equation}
    \hat{\mathbf{y}}_{\text{pos}} = f_{\text{LOC}}(\mathbf{H}; \mathbf{\Theta}_{\text{LOC}}),
\end{equation}
where $f_{\text{LOC}}(\cdot)$ denotes the localization NN, $\mathbf{\Theta}_{\text{LOC}}$ represents its parameters, and $\hat{\mathbf{y}}_{\text{pos}}$ is the predicted three-dimensional user position vector. The mean squared error (MSE) loss, commonly used in CSI-based localization tasks, defines the training objective as 
\begin{equation}
    \min_{\mathbf{\Theta}_{\text{LOC}}} \| \mathbf{y}_{\text{pos}} - \hat{\mathbf{y}}_{\text{pos}} \|^2,
\end{equation}
where $\mathbf{y}_{\text{pos}}$ is the ground-truth user position vector.
 
\subsubsection{Summative Synopsis}
In both of the sensing tasks described above, the accuracy of inference relies heavily on two key components: the NN models (${\rm f}_{\text{HAR}}(\cdot)$ and ${\rm f}_{\text{LOC}}(\cdot)$) and their associated parameters ($\mathbf{\Theta}_{\text{HAR}}$ and $\mathbf{\Theta}_{\text{LOC}}$), when trained on a given dataset. Existing studies have primarily focused on carefully designing NN architectures by drawing on successful practices from the CV domain (as shown in Table \ref{csi_cv_methods}), while often overlooking the potential of inheriting pre-trained parameters from CV models.

\section{The LVM4CSI Framework and its Workflow}
\label{s3}
Despite extensive efforts \cite{10582829,cui2025exploring,zheng2025beamllm,zhang2025port,zheng2024large} to improve CSI acquisition and utilization by fine-tuning the NN parameters of LLMs, such as GPT-2 and LLaMA 2, the potential of LVMs designed for CV data in wireless channels remains unexplored. In this section, we first present the motivation for directly applying LVMs to wireless channels. Then, the LVM4CSI framework and its corresponding workflow are introduced.

\subsection{Motivation}
\label{s31}

As previously noted, given that wireless channels and CV data belong to entirely distinct modalities, current CSI-related works typically adopt only the architectures of advanced CV NNs while avoiding the inheritance of their parameters. This approach, however, introduces a contradiction. Specifically, CV NNs are carefully designed to leverage the inherent characteristics of CV data, such as spatial hierarchies and local patterns. When applied to wireless channel tasks, these CV NN architectures assume that CSI data shares CV-like properties, which conflicts with the common understanding that the modalities are entirely distinct.
For example, the stripe structure of CSI in the angular-delay domain, resulting from the windowing effect in the discrete Fourier transform (DFT), is observed in \cite{10229094}. Consequently, a stripe-aware NN model from CV \cite{dong2022cswin} is adopted to enhance CSI acquisition accuracy. This highlights the contradiction and how CV NN architectures implicitly assume modality similarities. Therefore, since existing works have leveraged CV-like properties, it is worth considering the inheritance of CV NN parameters, which are closely tied to the characteristics of CV data.

On the other hand, as noted in Section \ref{s1}, the CSI image in Fig.~\ref{CSIimage} appears visually similar to the starry sky image (e.g., Fig.~\ref{star1911OD}) in Fig.~\ref{star1911} at first glance. Based on this observation, we directly apply the CV object detection NN DINO-X \cite{ren2024dino} to extract paths from the CSI image, in contrast to prior works \cite{9120709, 10345484} that train specialized NNs for this communication task. The detected signal paths, enclosed in purple boxes, are shown in Fig.~\ref{CSIimageOD} with promising results. DINO-X accurately detects all four paths, demonstrating the potential of applying LVMs to wireless channels. This strong performance can be attributed to the extensive data used in LVM training. As noted in \cite{ren2024dino}, DINO-X was trained on over 100 million image samples, enabling it to perform well on CV-like tasks.
Moreover, star detection in Fig.~\ref{star1911OD} performs significantly worse than path detection in the CSI image, as some stars remain undetected. This suggests that CSI-related tasks may often be simpler than CV tasks, mainly due to the sparse nature of wireless channels in most scenarios. Such simplicity further highlights the potential of applying LVMs to wireless channels.

\begin{figure}[t]
    \centering
    \includegraphics[width=0.5\textwidth]{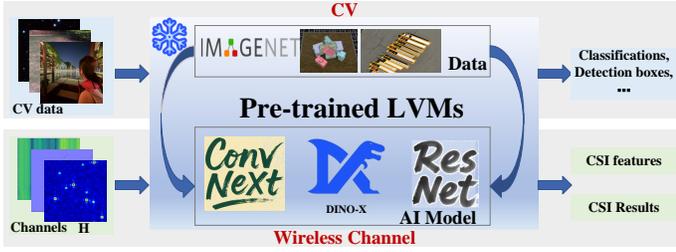} 
    \caption{\label{LVM4CSI}Key framework of the proposed LVM4CSI approach, where the NN parameters of the pretrained LVMs are frozen and the LVMs are directly utilized to conduct CSI-related tasks or produce CSI features for subsequent tasks.} 
\end{figure}

\begin{figure*}[t]
    \centering
    \includegraphics[width=0.95\textwidth]{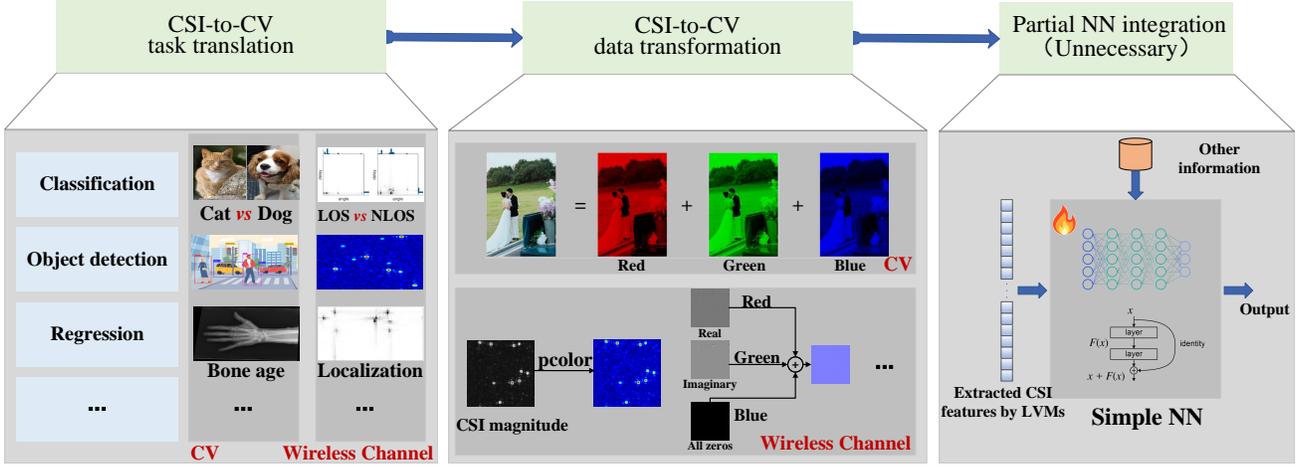}    \caption{\label{workflowLVM4CSI}Key workflow of the proposed LVM4CSI approach, including three key steps, namely CSI-to-CV task translation, CSI-to-CV data transformation, and partial NN integration.}
\end{figure*}

\subsection{Main Framework and Workflow}
Fig.~\ref{LVM4CSI} illustrates the general framework of the proposed LVM4CSI, which directly applies well pre-trained LVMs to handle complex CSI-related tasks. In CV tasks, CV data, primarily images, are processed by LVMs such as ConvNeXt and DINO-X to generate classification categories, detection bounding boxes, and more. Based on the discussion in Section \ref{s31}, wireless channels can be treated as CV data and processed by well pre-trained LVMs to either directly generate the final results of CSI-related tasks or extract high-level CSI features for subsequent task decisions, as shown in Fig.~\ref{LVM4CSI}.

The framework outlined above is straightforward and easy to understand. However, several critical issues must be addressed beforehand, as outlined below:
\begin{itemize}
    \item {\bf LVM selection:} With hundreds of well-established pre-trained LVMs available, identifying the most suitable model for CSI-related tasks presents a significant challenge. This decision is not trivial, as it requires careful consideration of multiple factors to ensure optimal performance in the context of wireless communication systems. 
    \item {\bf Modality adaptation:} Adapting an LVM pre-trained on CV data, such as images, to the unique nature of CSI data is another major challenge. This difficulty arises from the fundamental differences between CV and wireless data, which make it hard to directly input wireless channels into LVMs. Innovative approaches are needed to ensure seamless compatibility when applying LVMs to wireless channels.
\end{itemize}
To tackle the challenges outlined above, a comprehensive workflow is developed and explored for this framework. The operational workflow for the LVM4CSI framework is depicted in Fig.~\ref{workflowLVM4CSI}. It encompasses three key steps, namely CSI-to-CV task translation, CSI-to-CV data transformation, and partial NN integration, which are executed sequentially.

{\bf CSI-to-CV task translation:} CV and wireless are two completely different research fields with a significant gap between them. CV covers a wide variety of tasks across different areas. Taking image-related tasks as an example, these include classification (e.g., cat and dog classification), regression (e.g., bone age prediction), object detection (e.g., star detection), and more. Meanwhile, wireless channels also encompass a diverse array of tasks related to CSI acquisition and utilization. For instance, CSI acquisition includes tasks such as channel denoising, feedback, path extraction, and others. At first glance, the tasks in CV and wireless channels appear entirely distinct and cannot be readily analogized.

To select an LVM to support CSI-related tasks, the task correlation between CV and CSI tasks must be carefully analyzed, thereby facilitating the translation of CSI-related tasks into similar CV tasks. This process requires understanding the goals of CSI-related tasks and determining the key factors that influence performance, enabling a more accurate translation into CV tasks. For example, path extraction of CSI in the angular-delay domain can be likened to object detection, since both aim to determine object localization within an image. However, in a real CSI image, noise is inevitable and may be mistakenly identified as objects. Therefore, the object detection process must be capable of receiving instructions or prompts to detect the designated targets accordingly. Based on the preceding analysis, the path extraction task can be mapped to instruction-driven object detection in CV.

{\bf CSI-to-CV data transformation:} Once the CSI-related task is translated into a CV task and the appropriate LVM is selected, the focus shifts to ensuring that the CSI data can be effectively adapted to the selected LVM. This adaptation depends on two primary factors: the type of CSI image that the LVM can handle and the input format preferred by the LVM.

The type of CSI image selected is determined by the nature of the CSI-related tasks. For instance, the CSI image presented in Fig.~\ref{CSIimage} is represented in the angular-delay domain \cite{9120709, 10345484}, thereby visually depicting signal paths to facilitate path extraction. If the CSI is represented in the spatial-frequency domain, the signal path cannot be extracted, regardless of the methods employed, including the NOMP algorithm \cite{7491265} or object detection LVMs. Therefore, the CSI must be transformed into a domain or type where the CSI-related tasks can be effectively addressed.

CV data, such as images, is typically structured in a three-channel RGB format. In contrast, CSI data consists of complex values, necessitating its conversion into a three-channel RGB format to match the expected input of LVMs. For CSI-related tasks that depend solely on the element-wise modulus matrix of complex CSI in the angular-delay domain, the element-wise modulus matrix can be directly converted into a three-channel RGB format by generating a pseudocolor image or by representing the grayscale CSI modulus image in the RGB format. Conversely, for CSI-related tasks that utilize both the real and imaginary parts of CSI, an additional matrix can be incorporated as the third channel to create a three-channel CSI image. As illustrated in the middle part of Fig.~\ref{workflowLVM4CSI}, a simple all-zero matrix can be used as the third channel, thereby transforming the two-channel CSI image into a three-channel format without introducing additional information. While more advanced approaches could significantly enhance the performance of the proposed LVM4CSI framework, a simpler method is adopted here for illustrative purposes.

{\bf Partial NN integration:} In certain CSI-related tasks, LVMs can directly generate the desired final output. However, given that CV and wireless represent two fundamentally distinct domains, the final output of the LVM is often inapplicable to most CSI-related tasks. For instance, when a classification LVM is employed, the generated class output lacks relevance to wireless channels, rendering it meaningless in that context. Consequently, in most cases, LVMs are employed to extract features from CSI images for subsequent use in CSI-related tasks, with an additional partial NN integrated.

The core framework for integrating the additional NN is depicted in the right part of Fig.~\ref{workflowLVM4CSI}. Beyond the CSI data itself, other information may be essential or could further improve task performance. Moreover, certain information within the CSI may be lost during the CSI-to-CV transformation process. For instance, when the CSI matrix is normalized for three-channel image generation, power information may be lost, significantly impacting localization accuracy. Consequently, the additional partial NN requires two inputs: the CSI features extracted by the LVMs and the supplementary information deemed necessary. The output of this NN depends on the specific requirements of the CSI-related tasks. Given that the CSI features are high-level due to preprocessing by the LVMs, the partial NN performs only simple operations, and in some cases, a multilayer perceptron (MLP) with a single hidden layer suffices. This highlights an additional benefit of the proposed LVM4CSI framework: even when LVMs cannot directly produce the desired output, their application substantially simplifies the design and training requirements of the subsequent NN.

\section{Case Studies} 
\label{s4}
\subsection{Channel Estimation}
\label{s41}
According to the introduction in Section \ref{s2B}, the channel estimation problem in (\ref{pilotModel}) can be formulated to extract the three-tuple $\{\alpha, \Theta, T\} = \{\alpha_l, \Theta_l, T_l\}_{l=1,\ldots,L}$ from the noisy received pilot signal $\mathbf{Y}$, a matrix that can be regarded as a noisy channel \cite{9120709}. Since $L \ll M$, the noisy channel matrix $\mathbf{Y}$ becomes sparse when transformed from the spatial-frequency domain to the angular-delay domain using DFT operations.

In particular, following the approach in \cite{9120709}, the noisy channel matrix $\mathbf{Y}$ in the spatial-frequency domain is mapped to the angular-delay domain using sets of angle and delay bases, as expressed
\begin{equation}
\tilde{\mathbf{Y}} \in \mathbb{C}^{\beta M \times \gamma N} = \mathbf{F}_{\rm a}^T \mathbf{Y} \mathbf{F}_{\rm d}  ,
\label{eq:y_transform}
\end{equation}
where $\mathbf{F}_{\rm a} \in \mathbb{C}^{ M \times \beta M}$ and $\mathbf{F}_{\rm d} \in \mathbb{C}^{N \times \gamma N}$ represent the angle and delay bases, respectively. These are formed by taking the first $M$ rows of a $\beta M$-dimensional DFT matrix and the first $N$ rows of a $\gamma N$-dimensional DFT matrix, with oversampling factors $\beta, \gamma > 1$.

\begin{figure}[t]
    \centering
    \includegraphics[width=0.5\textwidth]{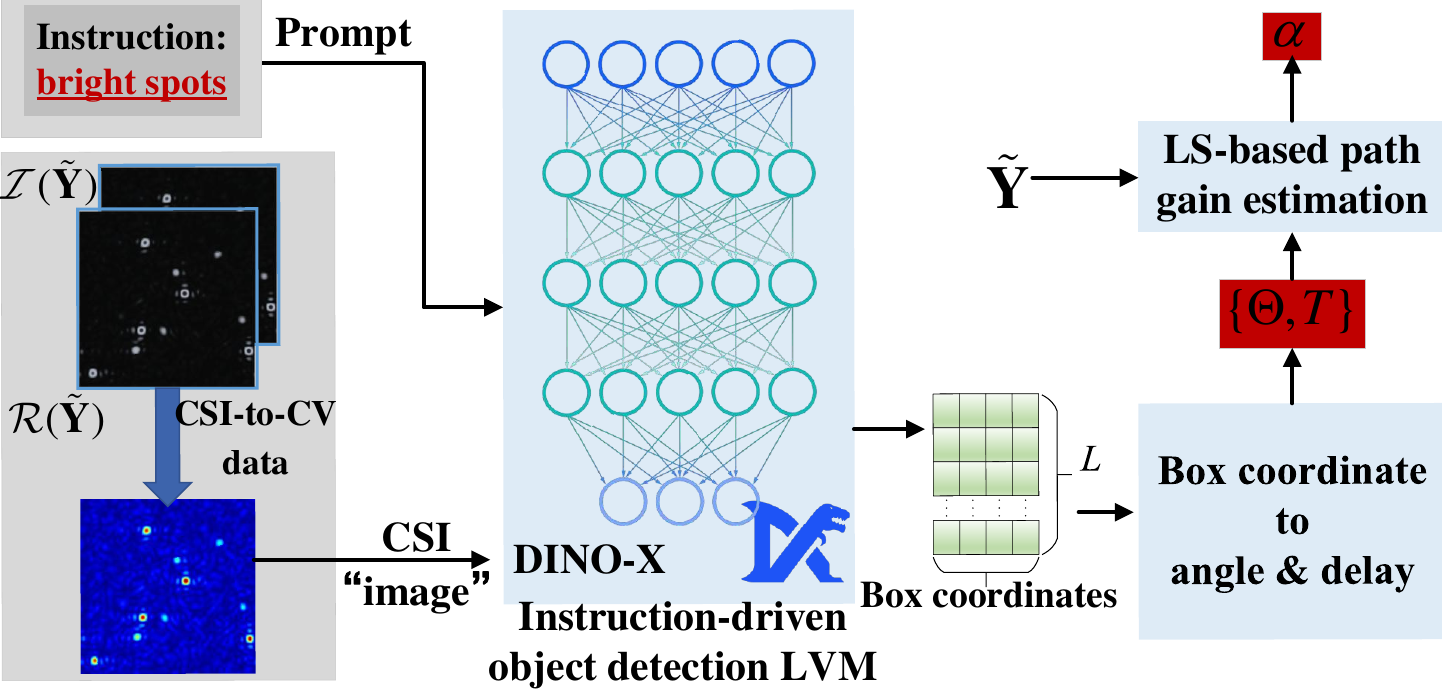}    \caption{\label{pathExtraction}Key framework of the LVM4CSI-enabled channel estimation, where the pre-trained DINO-X is directly utilized to path information with the prompt aid.}
\end{figure}

Some iterative algorithms, such as NOMP \cite{7491265} with a computational complexity of $\mathcal{O}(LMN \log_2(MN))$\footnote{The oversampling factors, $\beta$ and $\gamma$, are omitted here.}, extract the path information $\{\Theta, T\}$ from the sparse channel $\tilde{\mathbf{Y}}$, and then compute the path gains $\alpha$ via least squares (LS) algorithms. However, these iterative algorithms are computationally intensive and fail to satisfy the real-time requirements for CSI acquisition. In this section, as depicted in Fig.~\ref{pathExtraction}, we address this challenge by leveraging the proposed LVM4CSI framework, with the key steps outlined below.

Following the workflow illustrated in Fig.~\ref{workflowLVM4CSI}, the first step of the LVM4CSI framework involves translating the path extraction task into a CV task. As depicted in Fig.~\ref{CSIimage}, the signal paths in the noisy CSI image $\tilde{\mathbf{Y}}$ resemble bright spots against a dark background. The goal of path extraction is to identify the center position of each spot. This operation is analogous to object detection tasks in CV, where the bright spots in $\tilde{\mathbf{Y}}$ are treated as objects (signal paths), and their center positions, which map directly to the path parameters (angles and delays), are determined.

However, beyond the four brightest ``spots'' in the CSI image shown in Fig.~\ref{CSIimage}, additional information such as ``ripples'' resulting from DFT operations \cite{10229094} is also present. If the CSI image is directly processed by the object detection LVM, the unified model may inadvertently detect undesired objects. Therefore, the object detection process requires guidance through an instruction. Consequently, instruction-driven object detection LVMs, such as DINO-X, offer a promising solution. In these models, as depicted in Fig.~\ref{pathExtraction}, both the image and an instruction (prompt) that defines the specific objects to be detected are provided as inputs to the network. The outputs of the DINO-X LVM are the bounding box coordinates of the detected paths.

Assuming that $\hat{L}$ paths have been detected and the center coordinates of their bounding boxes are given by $\{w_l, h_l\}_{l=1,\ldots,\hat L}$, the angle and delay of the $l$-th path, denoted as $\hat{\Theta}_l$ and $\hat{T}_l$ respectively, can be calculated as 
\begin{equation}
    \hat{\Theta}_l = 1 - \frac{w_l}{\beta M},  \quad 
    \hat{T}_l =  \frac{h_l}{\gamma N}. 
\end{equation}
Once $\{\hat{\Theta}, \hat{T}\} = \{\hat{\Theta}_l, \hat{T}_l\}_{l=1,\ldots,\hat L}$ is obtained, the corresponding path gain $\hat{\alpha}$ can be calculated by the LS algorithm, and the final channel matrix $\hat{\mathbf{H}}$ can be reconstructed as 
\begin{equation}
\label{HyoloRec}
\hat{\bf H} = \sum_{l=1}^{\hat L} \hat\alpha_l \mathbf{a}(\hat\Theta_l) \mathbf{b}(\hat T_l)^{\text{T}}.
\end{equation}
The quality of the reconstructed CSI matrix $\hat{\mathbf{H}}$ depends on four key factors: the number of detected paths ($\hat{L}$), the accuracy of the bounding box centers ($\{\hat{\Theta}, \hat{T}\}$), the performance of the LS algorithm ($\hat{\alpha}$), and the noise power ($\sigma^2$).
 
In addition, as mentioned in the second step of the LVM4CSI workflow shown in Fig.~\ref{workflowLVM4CSI}, the CSI matrix should first be translated into CV data that can be processed by DINO-X. Since only the path centers need to be detected, the phase information of elements within the complex CSI matrix $\tilde{\mathbf{Y}}$ is unnecessary. Only the element-wise modulus matrix of the noisy CSI in the angular-delay domain, defined as $|\tilde{\mathbf{Y}}| \in \mathbb{R}^{\beta M \times \gamma N}$, is required. Therefore, only the matrix $|\tilde{\mathbf{Y}}|$ is used during path extraction. The matrix $|\tilde{\mathbf{Y}}|$ is first normalized within the range $[0,1]$, and then converted into a three-channel RGB image using the \texttt{jet} function in MATLAB.

\subsection{CSI-based Human Activity Recognition}

As discussed in Section \ref{HARmodel}, CSI-based human activity recognition can be treated as a classification task, which focuses on identifying activity categories from a sequence of CSI matrices, as formulated in (\ref{harmodelF}). In CV, numerous classification LVMs, such as ResNet \cite{he2016deep} and ConvNeXt \cite{Liu_2022_CVPR}, are available. However, these models are typically trained on natural image datasets like ImageNet, which labels 1,000 categories (e.g., barber chair, beer bottle, volleyball) that are unrelated to CSI-based activity recognition. Consequently, these CV-trained LVMs cannot be directly applied to classify human activity categories in this context. 

\begin{figure}[t]
    \centering
    \includegraphics[width=0.5\textwidth]{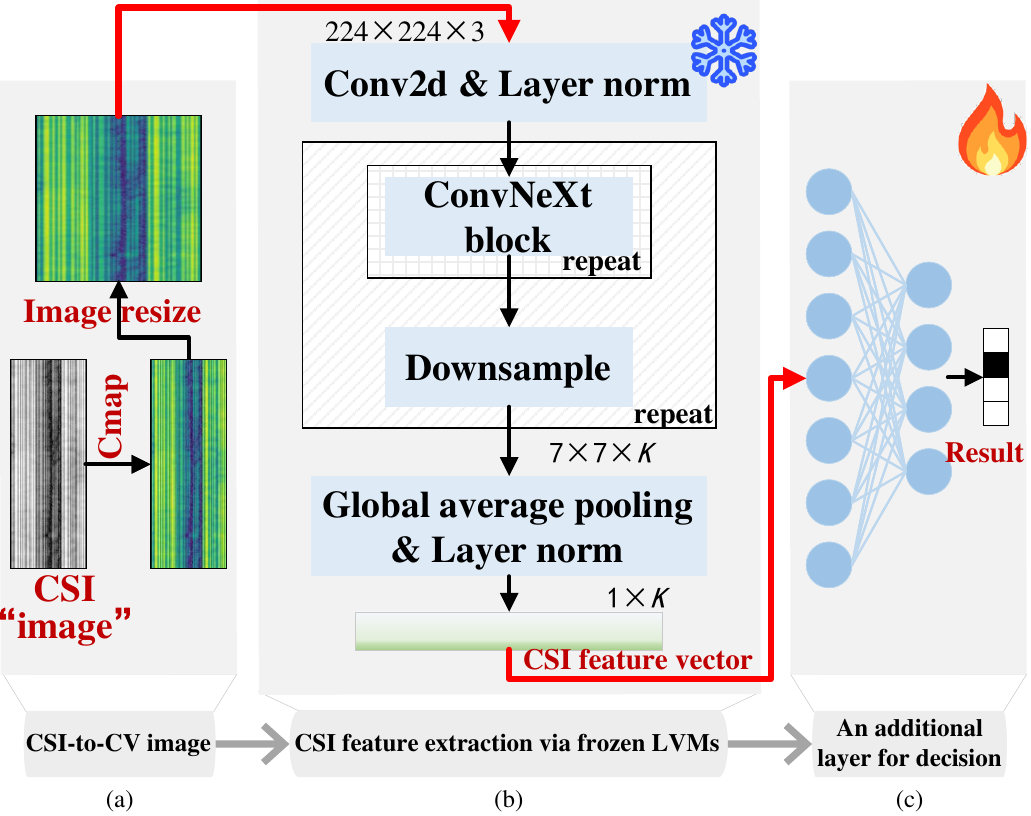} 
    \caption{\label{HARframework}Key framework of the LVM4CSI-enabled CSI-based human activity recognition, which utilizes the pre-trained frozen classification LVM, ConvNeXt, to extract CSI features, followed by a straightforward dense layer.} 
\end{figure}

Based on the preceding analysis, classification LVMs in this task are not employed for direct predictions but instead for extracting CSI features to facilitate subsequent category decisions. Consequently, as shown in the right part of Fig.~\ref{workflowLVM4CSI}, an additional NN must be integrated. The detailed framework for this approach, adhering to the guidelines outlined in Fig.~\ref{workflowLVM4CSI}, is illustrated in Fig.~\ref{HARframework}, comprising three key components:
\begin{itemize}
    \item {\bf CSI-to-CV image:} Human activity recognition is a coarse-grained task that categorizes CSI into several classes, such as the seven activity categories outlined in \cite{8067693}. Thus, only the element-wise modulus of the complex CSI matrix is used, directly discarding the phase information without considering the real and imaginary components separately. In this context, the input CSI image has a dimension of $T \times M \times N$, which is then reshaped into a grayscale image format of $T \times MN$, as depicted in the bottom-left of Fig.~\ref{HARframework}(a). Subsequently, the grayscale image is converted into an RGB three-channel image and resized to match the input dimension required by the classification LVM.
    
    \item {\bf CSI feature extraction via frozen LVMs:} Most classification LVMs consist of two primary modules, with the first used for image feature extraction and the second for class decision-making. As shown in Fig.~\ref{HARframework}(b), using the well-known ConvNeXt classification LVM as an example, the first module is utilized to extract features from the CSI image, yielding a feature vector of dimension \(1 \times K\).
    
    \item {\bf An additional layer for decision:} The feature vector with \(K\) elements generated in the preceding step is used to determine the activity category. Here, we adopt a single dense layer with softmax activation, designed to ensure minimal computational complexity.\footnote{This component can also be implemented using non-NN methods, such as the k-nearest neighbors (k-NN) algorithm. However, in this study, we directly employ a dense layer for simplicity.}
\end{itemize}
In contrast to the LVM4CSI-based channel estimation shown in Fig.~\ref{pathExtraction}, which requires no training, the additional dense layer here necessitates training. To accelerate this process, CSI feature vectors are first generated, followed by training of the dense layer using the loss function in (\ref{lossHAR}). This training process can be completed in an exceptionally short time owing to the minimal complexity of the dense layer.

\subsection{CSI-based User Localization}
\label{locaMethod}

Similar to the CSI-based human activity recognition discussed earlier, CSI-based user localization represents another key sensing task that leverages CSI data. Therefore, the core framework of LVM4CSI-based user localization closely resembles that of LVM4CSI-based human activity recognition, employing frozen LVMs to extract CSI features. However, user localization is a finer-grained regression problem, in contrast to the coarser classification nature of activity recognition. Consequently, more detailed CSI data, including phase information as highlighted in \cite{9129430}, must be incorporated for user localization. Specifically, both the real and imaginary components should be utilized.

In both the channel estimation and CSI-based human activity recognition tasks discussed earlier, the CSI image is first normalized before being processed by the LVMs, as channel power does not influence the path angle, delay, or activity classes. Even when these CSI images are not pre-normalized, as illustrated in Fig.~\ref{HARframework}(b), the LVM initially applies a normalization layer followed by a convolutional layer, effectively discarding the power information. However, unlike the aforementioned tasks, channel power is crucial for user localization, as it encapsulates essential information for accurate positioning. For instance, in a single-ray line-of-sight (LOS) channel scenario, the channel power directly corresponds to the path loss, which is determined by the distance between the BS and the user, that is, the user position.

\begin{figure}[t]
    \centering    \includegraphics[width=0.5\textwidth]{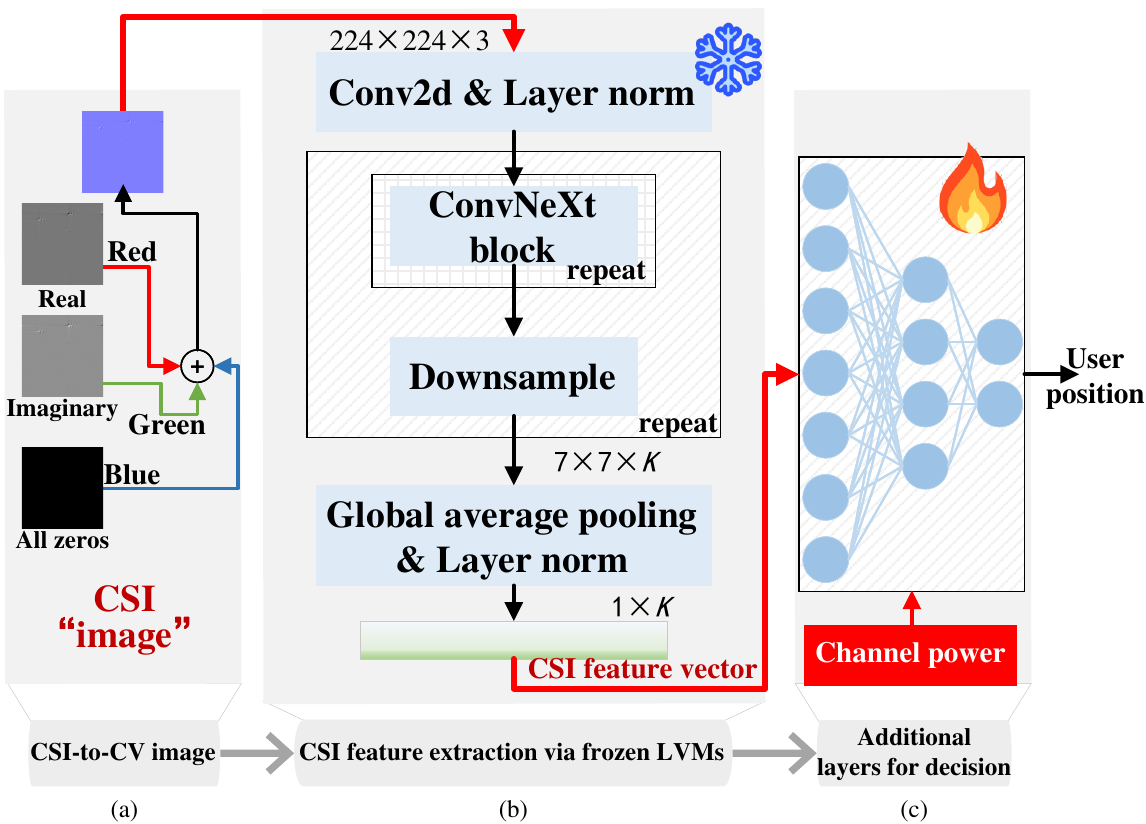} 
    \caption{\label{LocFramework}Key framework of the LVM4CSI-enabled CSI-based user localization, which utilizes the pre-trained frozen classification LVM, ConvNeXt, to extract CSI features. Several simple dense layers then determine the user position based on the extracted CSI features and channel power.} 
\end{figure}
 
The two aspects mentioned above necessitate careful consideration and meticulous design within the LVM4CSI-enabled CSI-based user localization framework. Fig.~\ref{LocFramework} presents the detailed architecture of this localization framework. Initially, both the real and imaginary components of the complex values are input into the NNs. To comply with the LVMs' input requirement of RGB three-channel images, a simple all-zero matrix is employed as the third channel, converting the normalized two-channel CSI image into a three-channel format without adding extra information, as depicted in the left part of Fig.~\ref{LocFramework}.

Subsequently, this three-channel image is processed by the LVM to extract CSI features, mirroring the process used in the human activity recognition task. This LVM-based feature extraction yields a $1 \times K$ feature vector. Unlike human activity recognition, which directly categorizes activity classes using the feature vector, localization is a fine-grained regression problem that also requires channel power, a single real value. Thus, the channel power is expanded into a $1 \times 8$ vector via a dense layer and then concatenated with the CSI feature vector, forming a combined vector of $1 \times (K+8)$. This concatenated vector is passed through three dense layers with 32, 16, and 2 neurons, respectively. The final dense layer, with 2 neurons and a Sigmoid activation function, outputs the user's two-dimensional position coordinates, while the preceding dense layers employ ReLU activation functions. Through these four simple dense layers, the extracted CSI feature vector is effectively integrated with the channel power, enabling accurate prediction of the user position.
 
\section{Numerical Results and Discussions}
\label{s5}
 
\subsection{Simulation Settings}

This work examines three different cases: channel estimation, CSI-based human activity recognition, and CSI-based user localization. The LVM-enabled frameworks for the first and last tasks can be trained and evaluated using simulated CSI data. However, accurately modeling human activity recognition in simulated environments is challenging, and thus, we utilize real CSI measurements obtained from practical systems for this task. 

\subsubsection{Channel estimation}
In this task, CSI data is generated based on (\ref{channelModel}). The number of transmit antennas \( M \) and subcarriers \( N \) are both set to 64, while the path number \( L \) varies as 2, 4, 6, 8, and 10. The channel estimation signal-to-noise ratio (SNR) ranges from 0 dB to 10 dB. The oversampling factors \( \beta \) and \( \gamma \) are both set to 4, resulting in the dimension of the processed CSI image in (\ref{eq:y_transform}) being \( 256 \times 256 \).
Unlike existing studies \cite{9120709, 10345484}, which generate large datasets for training path extraction NNs (e.g., 5,000 CSI samples in \cite{10345484}), the path extraction within the LVM4CSI-enabled framework directly utilizes well-trained LVMs and requires no training, thereby completely removing the need for NN design and dataset collection. Therefore, we generate only 50 CSI samples per setting to evaluate the proposed framework, following the workflow outlined in Fig.~\ref{workflowLVM4CSI}.

\subsubsection{CSI-based human activity recognition}
\label{HARcaseSet}
Since accurately modeling human activity recognition in simulated environments poses significant challenges, we employ practical CSI measurements from the public dataset in \cite{8067693}.\footnote{As no CSI datasets are available for cellular communications, we use this WiFi dataset instead.} Seven different human activities, namely lie down, fall, walk, pick up, run, sit down, and stand up, are considered in the practical measurement system.
A wireless router configured with three antennas (i.e., $M=3$) functions as the BS (access point), while the user employs a commercial Intel 5300 network interface card to collect CSI samples at a sampling rate of 1 kHz. The system operates in the 2.4 GHz frequency band with 30 subcarriers (i.e., $N=30$). Since human activities occur over continuous time periods rather than at an instant, the CSI extracted from 250 consecutive packets is treated as one CSI group, corresponding to $T=250$ in (\ref{harmodelF}). Therefore, the resulting CSI has a dimension of $250 \times 3 \times 30$. The training and test datasets contain 3,977 and 996 samples, respectively.

Since human activity recognition is a coarse-grained sensing application, we utilize only the element-wise modulus of the CSI group matrix. The original $250 \times 90$ real-valued matrix is subsequently transformed into a $224 \times 224$ RGB 3-channel image through color mapping and resizing operations. This preprocessing step ensures compatibility with the input dimension required by most classification LVMs.

The trainable NN in this task, as depicted in the right part of Fig.~\ref{HARframework}, comprises only a simple dense layer. The training process employs the Adam optimizer, with the training epoch, batch size, and learning rate set to 256, 200, and 0.001, respectively. The runtime for each epoch is approximately 140 ms, enabling the NN training to be completed within 30 seconds.

\begin{figure*}[!t]
        \centering
        \subfigure[NMSE vs. SNR]{\label{NMSEsnr}
\includegraphics[width=0.49\textwidth]{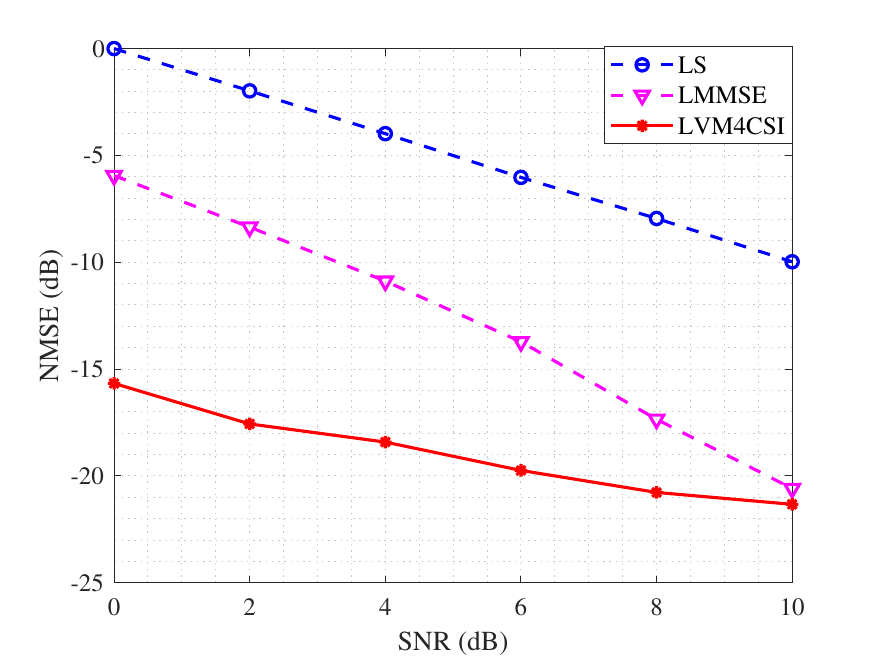}}
        \hfill
        \subfigure[NMSE vs. Path number]{\label{NMSEpath}     \includegraphics[width=0.49\textwidth]{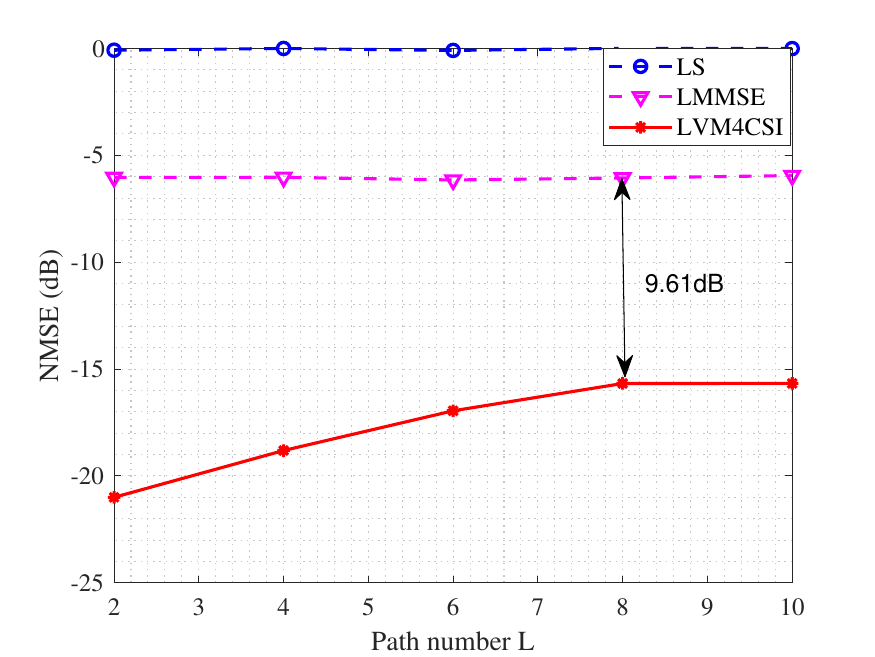}} 
        \caption{NMSE performance comparison of LVM-enabled channel estimation with LS and LMMSE methods across varying SNRs and path numbers. (a) The path number $L$ is set to 10. (b) The SNR is set to 0 dB.} 
\end{figure*}

\subsubsection{CSI-based user localization}
Unlike channel estimation, which requires only CSI samples, user localization necessitates both CSI samples and their corresponding ground-truth position labels. To generate these data pairs, we utilize the QUAsi Deterministic RadIo channel GenerAtor (QuaDRiGa) software,\footnote{\url{https://quadriga-channel-model.de/}} which produces channels along with associated positions using a statistical ray-tracing model.
The 3GPP\_38.901\_UMi\_NLOS channel model is adopted, with the path number \( L \) set to 10, and the center frequency and bandwidth configured at 2 GHz and 10 MHz, respectively. The number of transmitting antennas at the BS and the number of subcarriers are both set to 56. The BS is positioned at (0 m, 0 m, 25 m), while users are randomly distributed within a circular area centered at (200 m, 50 m, 1.5 m) with a radius of 50 m. The height of all users is uniformly set to 1.5 m, allowing the predicted user position to be represented as a two-dimensional vector. A total of 100,000 CSI samples and their corresponding positions are generated and stored, with 90 percent allocated to the training set and 10 percent to the test set.

The generated CSI data is transformed into the angular-delay domain according to (\ref{eq:y_transform}) with upsampling factors of 4. The resulting \(224 \times 224\) complex-valued upsampled CSI matrix is then converted into an RGB 3-channel image, following the procedure outlined in the left part of Fig.~\ref{LocFramework}. For the NN training process, the Adam optimizer is employed, with the training epoch, batch size, and learning rate set to 128, 200, and 0.001, respectively. Each epoch has an approximate runtime of 4 seconds.

\subsection{Benchmarks and Evaluation Metrics}

\subsubsection{Channel estimation}
In this task, the proposed LVM4CSI-enabled framework is evaluated against two conventional channel estimation methods: least squares (LS) and linear minimum mean square error (LMMSE). The performance is assessed using the normalized mean square error (NMSE) between the true CSI \(\mathbf{H}\) and the estimated CSI \(\hat{\mathbf{H}}\). The NMSE is defined as $\text{NMSE} = \mathbb{E}\{  {\|\mathbf{H} - \hat{\mathbf{H}}\|_2^2}/{  \|\mathbf{H}\|_2^2} \} $.

\subsubsection{CSI-based human activity recognition}
This task is a standard classification problem, and thus several well-established classification NNs\footnote{Although specialized NNs for CSI-based activity recognition may achieve higher accuracy, this study emphasizes the potential of applying LVMs to wireless channels rather than pursuing maximal accuracy. Therefore, standard NNs are selected for comparison.} are employed for comparative analysis, as follows: 
\begin{itemize}
    \item {\bf VGG19:} A 19-layer CNN using stacked $3\times 3$ filters to capture complex features. Its simple yet effective design enables robust image classification, and it has been widely applied in wireless channel analysis \cite{bu2022deep}.
    
    \item {\bf ResNet50:} A 50-layer residual NN with skip connections that facilitate stable deep learning. Its robust architecture allows for accurate classification of complex wireless signals \cite{bu2022deep}.
\end{itemize}

The final layer of both VGG19 and ResNet50 consists of a dense layer with 7 neurons, adjusted to match the number of human activity classes. Both models are trained from scratch, using the same hyperparameters specified in Section \ref{HARcaseSet}. Each training epoch takes approximately 6 seconds for VGG19 and 4 seconds for ResNet50, which is significantly longer than the single-dense-layer training used in the proposed LVM4CSI-enabled framework. The evaluation metric for this task is classification accuracy across the entire dataset.

\subsubsection{CSI-based user localization}
In this task, similar to the CSI-based human activity recognition case, we attempted to train established NN models such as VGG19 and ResNet50 from scratch. However, the training failed because the loss could not be reduced. Consequently, we design two specific NN models (one simple and one more complex) for this task to enable a fair comparison.

Since the upsampling operation in (\ref{eq:y_transform}) is applied to adjust the CSI matrix dimensions for LVM input, the benchmark NNs do not perform upsampling. Instead, the input dimension is set to $56 \times 56 \times 2$, corresponding to the number of antennas, subcarriers, and the two real and imaginary channels.
 
\textbf{NoFeatExt-CSI:} The input $56 \times 56 \times 2$ matrix is flattened into a vector and normalized. This vector, along with the channel power, is then fed into the additional NN module as shown in the right part of Fig.~\ref{LocFramework}. Compared with the proposed LVM-enabled localization model, this benchmark deliberately omits CSI feature extraction to highlight the importance of feature extraction in localization. Therefore, this benchmark is named ``NoFeatExt-CSI.''

\textbf{ConvFeatExt-CSI:} The input $56 \times 56 \times 2$ matrix is passed through three convolutional layers with 8, 32, and 1024 filters, respectively. Each layer uses a $3 \times 3$ kernel, a stride of 2, same padding, and ReLU activation. The resulting output is globally average pooled and flattened into a vector, which serves as the extracted CSI feature. This vector, combined with the channel power, is fed into the additional NN modules, as illustrated in the right part of Fig.~\ref{LocFramework}. In contrast to the proposed localization method, which uses pre-trained LVMs for CSI feature extraction, this benchmark employs a custom-designed and trained CNN module specifically for CSI-based localization. Hence, it is referred to as ``ConvFeatExt-CSI.''

\subsection{Results and Discussions}

\subsubsection{Channel estimation}
 
\begin{figure}[!t]
    \centering
    \subfigure[Path detection result]{\label{output_bboxes}
        \includegraphics[width=0.23\textwidth]{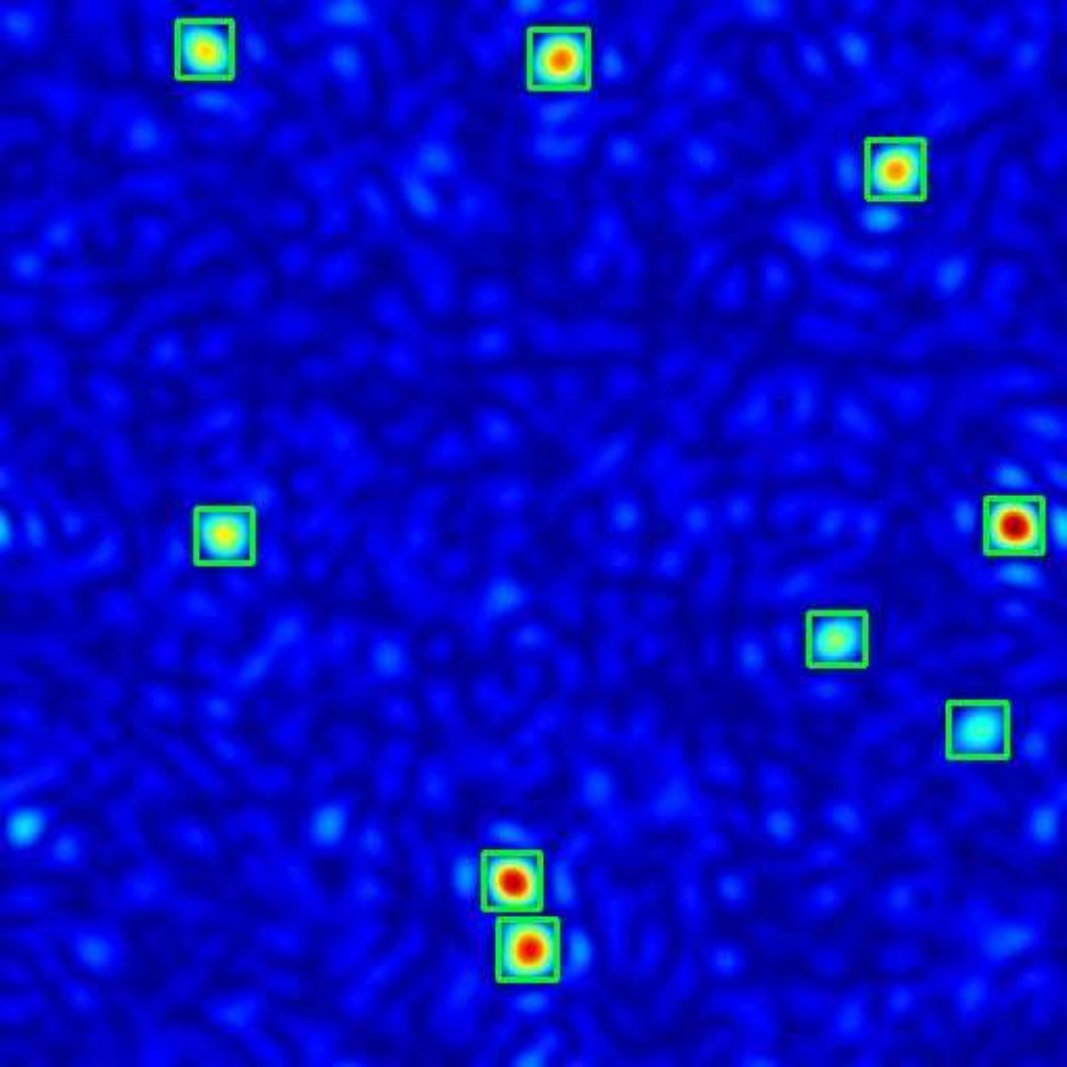}}
    \hfill
    \subfigure[Ground truth]{\label{output_ground_truth}
        \includegraphics[width=0.23\textwidth]{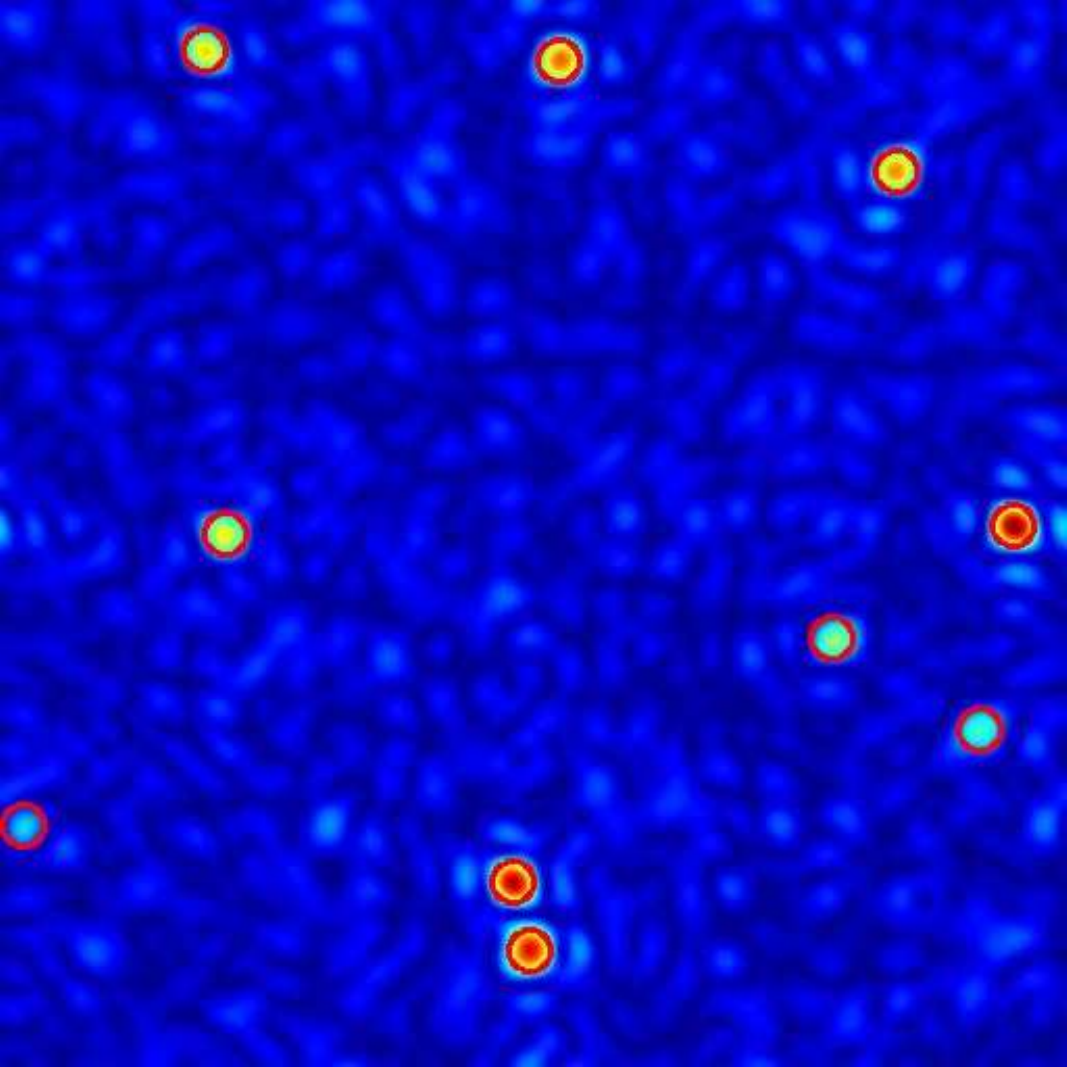}} 
    \caption{An example of low-quality channel estimation using the LVM4CSI framework, where the left bright spot (path) is submerged in noise and missed during detection.}
    \label{outputCsi}
\end{figure}

\begin{table*}[t]
  \centering
   \caption{Accuracy and Number of Trainable Parameters for LVM4CSI and Train-from-Scratch Models in CSI-Based Human Activity Recognition} 
  \label{tabAccuracy_comparison}
  \resizebox{\textwidth}{!}{ 
    \begin{tabular}{l *{9}{c}}
      \toprule
      & \multicolumn{6}{c}{LVM4CSI} & \multicolumn{3}{c}{Train-from-Scratch} \\
      \cmidrule(lr){2-7} \cmidrule(lr){8-10}
      Method & MobileNetV3 & VGG19  & ResNet50 & ConvNeXt (Tiny) & ConvNeXt (Base) & ConvNeXt (XLarge) & VGG19 & ResNet50 & Recurrent ConFormer \cite{10084426} \\
      \midrule
      Accuracy (\%) & 91.4 & 91.8 & 93.0 & 94.0 & 94.4 & \cellcolor{gray!30}95.8 & 92.8 & 94.0 & 96.2 \\
      \midrule
      Number of trainable NN parameters & 7,007  & 28,679  & 14,343  & 5,383  & 7,175 & \cellcolor{gray!30}14,343  & 139,598,919  & 23,548,935  & $\sim$500,000 \\
      \bottomrule
    \end{tabular}  }
\end{table*}

Fig.~\ref{NMSEsnr} illustrates the NMSE performance comparison of the proposed LVM4CSI-enabled channel estimation with LS and LMMSE methods across varying SNR levels, where the path number $L$ is set to 10. All three methods show improved performance as the SNR increases. As expected, the LMMSE method consistently outperforms the LS method. Notably, the LVM4CSI-enabled method outperforms both LS and LMMSE across all SNR values.
However, the performance advantage of LVM4CSI over LMMSE narrows at higher SNRs. For instance, the NMSE gap between LVM4CSI and LMMSE is 9.72 dB at 0 dB SNR, but only 0.74 dB at 10 dB SNR. At higher SNRs, channel estimation becomes less challenging, allowing LMMSE to perform comparably to LVM4CSI. In addition, existing learning-based channel estimation methods \cite{9127834} often perform well only within the SNR ranges they were trained on. In contrast, the LVM4CSI framework maintains robust performance across a wide SNR range, demonstrating its strong generalization ability enabled by pre-trained LVMs.

Fig.~\ref{NMSEpath} shows NMSE performance across different path numbers $L$ under a fixed SNR of 0 dB. LVM4CSI consistently outperforms both LS and LMMSE. The performance gap between LVM4CSI and LMMSE remains substantial, with the minimum NMSE gap reaching 9.61 dB. The NMSE of LS and LMMSE remains relatively stable across different path numbers due to fixed SNR and normalized power. In contrast, LVM4CSI performance depends on accurately detecting paths, which becomes more difficult as the number of paths increases.
Fig.~\ref{outputCsi} presents a failure case where one path is undetected because it is obscured by noise. In this example, with ${\rm SNR} = 0$ dB and $L = 10$, nine out of ten paths are correctly detected, and the resulting NMSE is $-13.31$ dB, which is above the average NMSE.

\subsubsection{CSI-based human activity recognition}

The LVM4CSI framework for CSI-based human activity recognition employs four classification LVMs: VGG19 \cite{vggnet}, MobileNetV3 \cite{Howard_2019_ICCV}, ResNet50 \cite{he2016deep}, and ConvNeXt \cite{Liu_2022_CVPR}. Three ConvNeXt variants, namely Tiny, Base, and XLarge, are evaluated. Additionally, a task-specific model, Recurrent ConFormer \cite{10084426}, is included for comparison. 

Table~\ref{tabAccuracy_comparison} summarizes classification accuracy and trainable parameter counts. 
\begin{itemize}
    \item In LVM4CSI, classification performance improves with increasing model complexity. MobileNetV3, designed for mobile devices, achieves the lowest accuracy due to its minimal complexity. Among the ConvNeXt variants, classification accuracy improves from 94.0\% (Tiny) to 95.8\% (XLarge) as model size increases.

    \item We also trained VGG19, ResNet50, and ConvNeXt from scratch. However, the ConvNeXt models failed to converge due to the small dataset. Therefore, only results for VGG19 and ResNet50 are included. These models outperform their frozen counterparts within the LVM4CSI framework. Still, they fall short compared to the pre-trained ConvNeXt models, showing the effectiveness of LVM pretraining.

    \item Although Recurrent ConFormer achieves the highest accuracy of 96.2\%, it requires significant design effort and hardware resources. In contrast, if LVMs like ConvNeXt are pre-trained on datasets that include some CSI, they can deliver competitive or superior performance with much less effort.
\end{itemize}
Although the LVM4CSI framework uses far fewer trainable parameters, approximately 1/30 of those in the Recurrent ConFormer, it still achieves comparable or even superior accuracy. This highlights the practicality and efficiency of applying pre-trained LVMs to CSI-related tasks.

\begin{figure}[t]
    \centering        \includegraphics[width=1\linewidth]{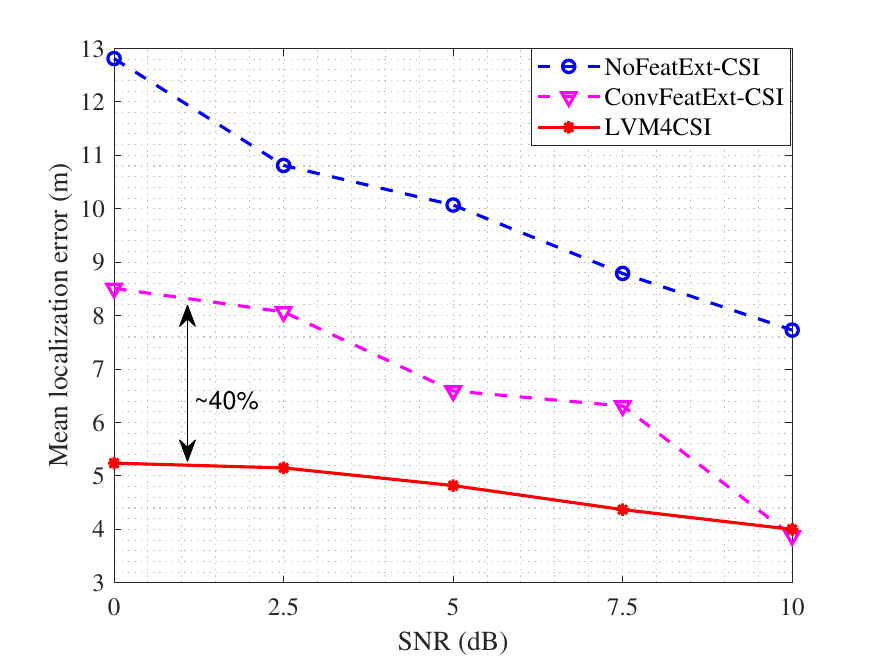}
        \caption{Mean localization error comparison between the LVM4CSI framework and two benchmarks. LVM4CSI reduces the localization error by approximately 40\% compared to ConvFeatExt-CSI.} 
        \label{LocResult}
\end{figure}

\subsubsection{CSI-based user localization}

Since user localization is more complex than human activity recognition, the LVM4CSI framework uses the ConvNeXt (XLarge) model. Fig.~\ref{LocResult} presents a comparison of mean localization errors for LVM4CSI and two benchmark models across varying SNRs. The number of trainable parameters in NoFeatExt-CSI, ConvFeatExt-CSI, and LVM4CSI is 214,114, 334,106, and 33,634, respectively.

ConvFeatExt-CSI significantly outperforms NoFeatExt-CSI. For example, at 2.5 dB SNR, ConvFeatExt-CSI achieves a mean error of 8.07 meters compared to 10.81 meters for NoFeatExt-CSI. This result highlights the importance of effective feature extraction.
LVM4CSI achieves the best performance across all SNRs, especially at low SNR. At 2.5 dB SNR, it reduces localization error by 2.92 meters (approximately 40\%) compared to ConvFeatExt-CSI. As SNR increases, the performance gap narrows. At high SNR, feature extraction becomes easier, and the advantages of pre-trained LVMs are less pronounced.

\section{Conclusion and Future Directions}
\label{s6}

This paper proposes LVM4CSI, a novel framework that directly employs pre-trained LVMs with frozen parameters for wireless channel acquisition and utilization. It eliminates the need for task-specific NN design or extensive training. By translating CSI-related tasks into equivalent CV problems, transforming CSI data into CV-compatible formats, and integrating minimal additional NN layers, LVM4CSI leverages the robust feature extraction capabilities of frozen LVMs.
Case studies on channel estimation, human activity recognition, and user localization demonstrate that the LVM4CSI framework outperforms conventional and train-from-scratch NNs. It improves channel estimation NMSE by more than 9.61 dB compared to LMMSE without additional NN design or training, achieves comparable human activity recognition accuracy with less than 1/30 of the trainable parameters required by state-of-the-art methods, and reduces localization error by 40\% compared to specialized NNs.

Although the LVM4CSI framework shows significant potential for wireless channel applications, several key challenges remain. First, it is important to develop unified LVMs capable of addressing a broad range of CSI-related tasks to streamline task translation. Second, careful strategies are needed to incorporate CSI data into LVM pre-training in order to align the model's learned knowledge with wireless channel characteristics. Third, optimizing LVM architectures for real-time processing on resource-constrained devices is critical for practical deployment. Finally, exploring hybrid approaches that combine LVMs with domain-specific NNs may help balance performance and computational efficiency.

\bibliographystyle{IEEEtran}
\bibliography{magazine}

\end{document}